\documentclass[preprint2]{aastex}

\newcommand{\etal}{\mbox{et~al.}}

\def\deg      {{\ifmmode^\circ\else$^\circ$\fi}} 

\def\multidrizzle{{\tt MultiDrizzle}}
\def\tinytim{{\tt TinyTim}}

 \shorttitle{COSMOS-PSF}
 \shortauthors{Rhodes et al.}
 
\begin{document}

 \title{The Stability of the Point Spread Function of the
 \emph{Advanced Camera for Surveys} on the \emph{Hubble Space Telescope} and
 Implications for Weak Gravitational Lensing}

 \author{
Jason D. Rhodes\altaffilmark{1,2},
Richard Massey\altaffilmark{2},
Justin Albert\altaffilmark{2},
Nicholas Collins\altaffilmark{3},\\
Richard S. Ellis\altaffilmark{2},
Catherine Heymans\altaffilmark{4},
Jonathan P. Gardner\altaffilmark{3},
Jean-Paul Kneib\altaffilmark{5},\\
Anton Koekemoer\altaffilmark{6},
Alexie Leauthaud \altaffilmark{5},
Yannick Mellier\altaffilmark{7,8},
Alexandre Refregier\altaffilmark{9}, \\
James E. Taylor\altaffilmark{2,10} \& 
Ludovic Van Waerbeke\altaffilmark{4}}


\altaffiltext{$\star$}{Based on observations with the NASA/ESA {\em Hubble Space Telescope}, obtained at the Space Telescope Science Institute, which is operated by AURA Inc, under NASA contract NAS 5-26555; also based on data collected at : the Subaru Telescope, which is operated by the National Astronomical Observatory of Japan; the XMM-Newton, an ESA science mission with instruments and contributions directly funded by ESA Member States and NASA; the European Southern Observatory, Chile; Kitt Peak National Observatory, Cerro Tololo Inter-American Observatory, and the National Optical Astronomy Observatory, which are operated by the Association of Universities for Research in Astronomy, Inc. (AURA) under cooperative agreement with the National Science Foundation; the National Radio Astronomy Observatory which is a facility of the National Science Foundation operated under cooperative agreement by Associated Universities, Inc ; and the Canada-France-Hawaii Telescope operated by the National Research Council of Canada, the Centre National de la Recherche Scientifique de France and the University of Hawaii. }

\altaffiltext{1}{Jet Propulsion Laboratory, Pasadena, CA 91109}
\altaffiltext{2}{California Institute of Technology, MC 105-24, 1200 East
California Boulevard, Pasadena, CA 91125}
\altaffiltext{3}{Exploration of the Universe Division, Observational Cosmology Laboratory, Code 665, Goddard Space Flight Center, Greenbelt, MD
20771}
\altaffiltext{4}{University of British Columbia, 6224
Agricultural Road, Vancouver, B.C. V6T 1Z1, Canada}
\altaffiltext{5}{Laboratoire d'Astrophysique de Marseille, BP 8, Traverse
du Siphon, 13376 Marseille Cedex 12, France}
\altaffiltext{6}{Space Telescope Science Institute, 3700 San Martin
Drive, Baltimore, MD 21218}
\altaffiltext{7}{Institut d'Astrophysique de Paris, UMR7095 CNRS,
   Universit\'e Pierre~\&~Marie Curie - Paris, 98 bis bd Arago, 75014
   Paris, France.}
\altaffiltext{8}{Observatoire de Paris, LERMA, 61, avenue de
   l'Observatoire, 75014 Paris, France.}

\altaffiltext{9}{Service d'Astrophysique, CEA/Saclay, 91191 Gif-sur-Yvette, France}
\altaffiltext{10}{Department of Physics and Astronomy, University of Waterloo, Waterloo, Quebec, Canada}

\begin{abstract}

We examine the spatial and temporal stability of the \emph{Hubble Space Telescope's Advanced Camera for Surveys (ACS)} Wide Field Camera (WFC) point spread function (PSF) using the two square degree COSMOS survey.  This is particularly important for studies of weak gravitational lensing, where the ability to deconvolve the PSF from galaxy shapes is of paramount importance.  We show that stochastic aliasing of the PSF necessarily occurs during `drizzling', the image reduction step where individual exposures are combined and the large geometric distortion in the ACS WFC is removed.  This aliasing is maximal if the output pixel scale is equal to the input pixel scale of 0.05''.  Using simulated stars, we show that this source of PSF variation can be significantly reduced by choosing a Gaussian drizzle kernel with a size of 0.8 input pixels and by setting the output pixel size to 0.03''.  This result has been tested mainly on the COSMOS images which have four input exposures each dithered by about 60 pixels in the \emph{y} direction. However,  the recommendation holds for any ACS WFC data set taken with a similar number of input exposures and a dither pattern that does not cause output pixels to be derived from  input pixels that are a significant fraction of the ACS WFC chips away from each other in different exposures.  We show that the PSF is temporally unstable, most likely due to thermal fluctuations in the telescope's focus.  We find that the primary manifestation of this thermal drift in COSMOS images is an overall slow periodic focus change.

Using a modified version of the \tinytim\ PSF modeling software, we create  undistorted stars in a $30\times30$ grid across the ACS  WFC CCDs.  These PSF models are created for telescope focus values in the range $-10\mu$m to $+5\mu$m at one micron increments, thus spanning the allowed range of telescope focus values.  We  then use the approximately ten well measured stars in each COSMOS field to pick the best-fit focus value for each field.  The \tinytim\ model stars are then used to perform PSF corrections for weak lensing allowing  systematics due to incorrectly modeled PSFs to be  greatly reduced by using this \tinytim\ method as compared to using stars within the COSMOS fields alone, as is typically done in weak lensing analyses. We have made the software for PSF modeling using our modified version of \tinytim\ available to the astronomical community.  We show   the effects of Charge Transfer Efficiency (CTE) degradation, which distorts the object in the readout direction, mimicking a weak lensing signal.  We derive a parametric correction  for the effect of CTE on the shapes of objects in the COSMOS field as a function of observation date, position within the ACS WFC field, and object flux.  Finally, we discuss future plans to improve the CTE correction.

\end{abstract}


 \keywords{instrumentation: detectors---  surveys ---techniques: image processing  }



 \section{Introduction}

The addition of the Advanced Camera for Surveys (ACS; Pavlovsky \etal 2006) to the  \emph{Hubble Space Telescope} (\emph{HST}) in February 2002 enabled significant science returns in a variety of disciplines due to the increased resolution, areal coverage, and quantum efficiency of ACS Wide Field Camera (WFC) as compared to previous HST imaging instruments including WFPC, WFPC2, and STIS.  Despite the myriad successes of ACS, some areas of study have been hampered by a lack of understanding of the properties of the ACS WFC Point Spread Function (PSF).  In particular, studies of weak gravitational lensing, in which the shapes of background galaxies undergo a small coherent distortion by foreground dark matter, are made difficult by  time variability of the ACS PSF.  Since we are interested in weak gravitational lensing, we concentrate our studies on the ACS WFC and do not study the PSF of other ACS channels.  Hereafter, when we refer to the ACS, we are referring to the ACS WFC.   Because the slight shape distortions due to weak lensing are up to an order of magnitude smaller than the shape distortions of small galaxies due to the ACS PSF, it is critical to have accurate PSF models for use in deconvolving the galaxy shapes from the PSF.   There are notable exceptions in which ACS has had great success in weak lensing, including the study of weak lensing by galaxy clusters (e.g Lombardi \etal\ 2005; Jee et al. 2005; Jee et al. 2006), and the results of Heymans \etal\ (2005) using the GEMS survey. In the case of lensing by galaxy clusters, the weak lensing signal is typically  large compared to the signal arising from large-scale structure alone and thus PSF effects are less important.   For clusters at $z>1$, PSF effects become important, and Jee \etal (2006) have found that a limited number of PSF models derived from globular cluster observations can be used for PSF correction.  The GEMS survey contained many images taken in a relatively short amount of time such that the PSF time variations could be internally calibrated using stars in the survey.  However, for surveys taken over an extended period of time in which there are few stars and the weak lensing signal is low, the time variability of the ACS PSF precludes using the stars in each exposure to make a PSF model suitable for weak lensing PSF deconvolution.  In this paper we offer an alternative solution in which templates of model PSFs are created and the few available stars in each ACS field are used to select the appropriate PSF template to be used for PSF deconvolution. A similar method using templates created by combining the stars from many exposures with similar PSF properties was used by Schrabback \etal\ (2006). Another alternative approach is to use globualr cluster fields with many stars to model the PSF (e.g. Jee \etal 2005; Jee \etal 2006). In addition to the temporally varying telescope PSF,  degradation of the charge transfer efficiency (CTE) of the ACS CCDs is gradually adding additional variation to the measured PSF.  The GEMS survey, taken relatively early in the life of ACS did not necessitate a correction for this effect but more recent surveys do need to take this effect into account.  This paper also addresses that problem by providing a prescription by which
measured galaxy shapes can be corrected for the effects of CTE degradation.

The Cosmic Evolution Survey (COSMOS) 2 square degree field ($\sim1.64$ square degrees with ACS; Scoville \etal\ 2006) has a unique combination of area, depth, and resolution that opens up new areas of study with the HST. This is particularly true in weak lensing where the surface density of resolved galaxies (up to $\sim80$ per square arcminute) is significantly higher than that available from ground-based surveys (10 to 30 to per square arcminute). Along with the excellent photometric redshifts available in the COSMOS field due to extensive ground-based follow-up, this allows studies of cosmic shear tomography at small angular scales to unprecedented accuracy  (Massey \etal\ 2006a).  COSMOS also uniquely affords the opportunity for high -resolution dark matter maps over a wide field (Massey \etal\ 2007). Such analyses depend critically on the ability to deconvolve the effects of PSF from the measured galaxy shapes and we demonstrate in this paper a procedure for doing so.  The demanding nature of our weak lensing analysis will provide PSF models of immediate applicability to other studies, including an examination of AGN in the COSMOS field (Gabor \etal\  2006).

Typical ground-based weak lensing surveys use the stars within each field to make a model of the PSF.  This PSF model, often represented as a polynomial across the  field, is then used to deconvolve the PSF from galaxy shapes using a variety of methods (see Heymans \etal\ 2006 and Massey \etal\ 2006b for a description of many of these methods). The modeling and deconvolution of the PSF remains the primary systematic for weak lensing studies and considerable work is currently going into optimizing the methods for doing this (again, see Heymans \etal 2006).  HST images useful for weak lensing contain too few stars in each image to make a viable PSF model.  In the past, this has been overcome by collecting stars from many images to model the PSF (e.g Rhodes, Refregier, \& Groth 2001; Hoekstra \etal\ 1998). However, we find that the ACS PSF is not sufficiently stable to allow for the collection of stars from many images that will provide models at the accuracy level we needed for cosmic shear.  Thus, we have modified the PSF creation software TinyTim (Krist \& Hook  2004) to create simulated PSFs that we use for PSF correction.  The reasons the ACS PSF is more temporally unstable than that of previous cameras such as WFPC2 and are two-fold. First, WFPC2 is  on-axis , while ACS is significantly off-axis yielding a more
sensitive measure of defocus.  Second, ACS has  a more
asymmetric distortion compared to  the largely radially symmetric
pattern for WFPC2. These effects combine to provide more sensitivity
to focus changes.

The degradation of CTE is caused by charge traps created by  charged particle  radiation damage induced CCD defects (Riess \& Mack 2004; Mutchler \& Sirianni 2005).  These traps cause a trailing of the charge during readout that can mimic a weak lensing signal.  This effect is particularly insidious for weak lensing applications because it preferentially affects low flux objects (such as galaxies) rather than high flux objects (such as bright stars).  So, the typical weak lensing method of using high signal-to-noise (S/N) stars to correct the PSF of low signal to noise galaxies completely fails in the case of CTE degradation.
Nor, strictly speaking, does it act as a convolution, although it has been treated as such to first order in the past (e.g. Rhodes \etal\ 2004).
The effect of CTE degradation depends in a complicated way on the date of observation, the position within the CCD, the flux of a source, and the background level of the image.
The optimal solution to the problems caused by CTE degradation would be to understand and remove the effects of CTE on the images themselves as the first step in image reduction.  However, the effects of CTE on individual pixel values has not yet been fully quantified and will be investigated in  a future paper.  We correct for CTE effects by subtracting a model of the spurious (CTE induced) signal from  the measured galaxy ellipticities at the end of our weak lensing pipeline.

This paper is organized as follows.  \S~\ref{dataandsoft} describes the data and software tools we use to analyze the data.  This section includes descriptions of the COSMOS survey and auxiliary data sets we use in \S~\ref{data}, a brief description of the multidrizzle image reduction pipeline in \S~\ref{multi},  a description of the TinyTim PSF simulation software in \S~\ref{tt}, and a description of our weak lensing PSF correction scheme in \S\ref{rrg}.  We detail the image reduction procedures we use for weak lensing images in \S~\ref{reduction}. In \S~\ref{models} we describe the TinyTim-based PSF models we have created and the procedure we use to deconvolve the   PSF from galaxy shapes is given in \S~\ref{correction}.  We conclude in \S~\ref{conclusions}, where we describe how the methods we have detailed can be of use on other data sets and with other weak lensing methods. In the conclusion we describe steps we have taken to make our PSF modelling software available to the community for use on other HST data sets.

\section{Data and Software}\label{dataandsoft}
\subsection{Data}\label{data}
The primary motivation for our analysis of the ACS PSF is to perform various weak gravitational lensing analyses on the COSMOS field.
The COSMOS field is a contiguous square covering 1.637 square degrees centered at 10:00:28.6, +02:12:21.0 (J2000) COSMOS was imaged with the ACS Wide Field Camera (WFC) during during HST cycles 12 (proposal ID 9822) and 13 (proposal ID 10092), between October 2003 and June
2005.  This data set  contains 575 slightly overlapping pointings, each with four individual exposures of 507 seconds dithered by (5,60) pixels.  At this depth, each  10 square arcminute  COSMOS pointing contains approximately 800 galaxies useful for weak lensing (sufficient size and signal-to-noise (S/N) and approximately 10 stars suitable for measuring the PSF.
For a more complete description of the COSMOS field see Scoville \etal 2006.
COSMOS was imaged in the F814W (\emph{I}) filter and our measurements of PSF are specific to that filter.  However, the methods we have developed are general and can be applied to other ACS WFC filters using the software we have made public (see \S\ref{conclusions}).  The applicability of our software and methods is limited by the ability of our modelling software, which does not include such effects as the long wavelength (e.g. red) scattering within the ACS WFC CCDs (see for instance, Sirianni \etal 2005).

\begin{figure}[h!]
\epsscale{1.05}
\plottwo{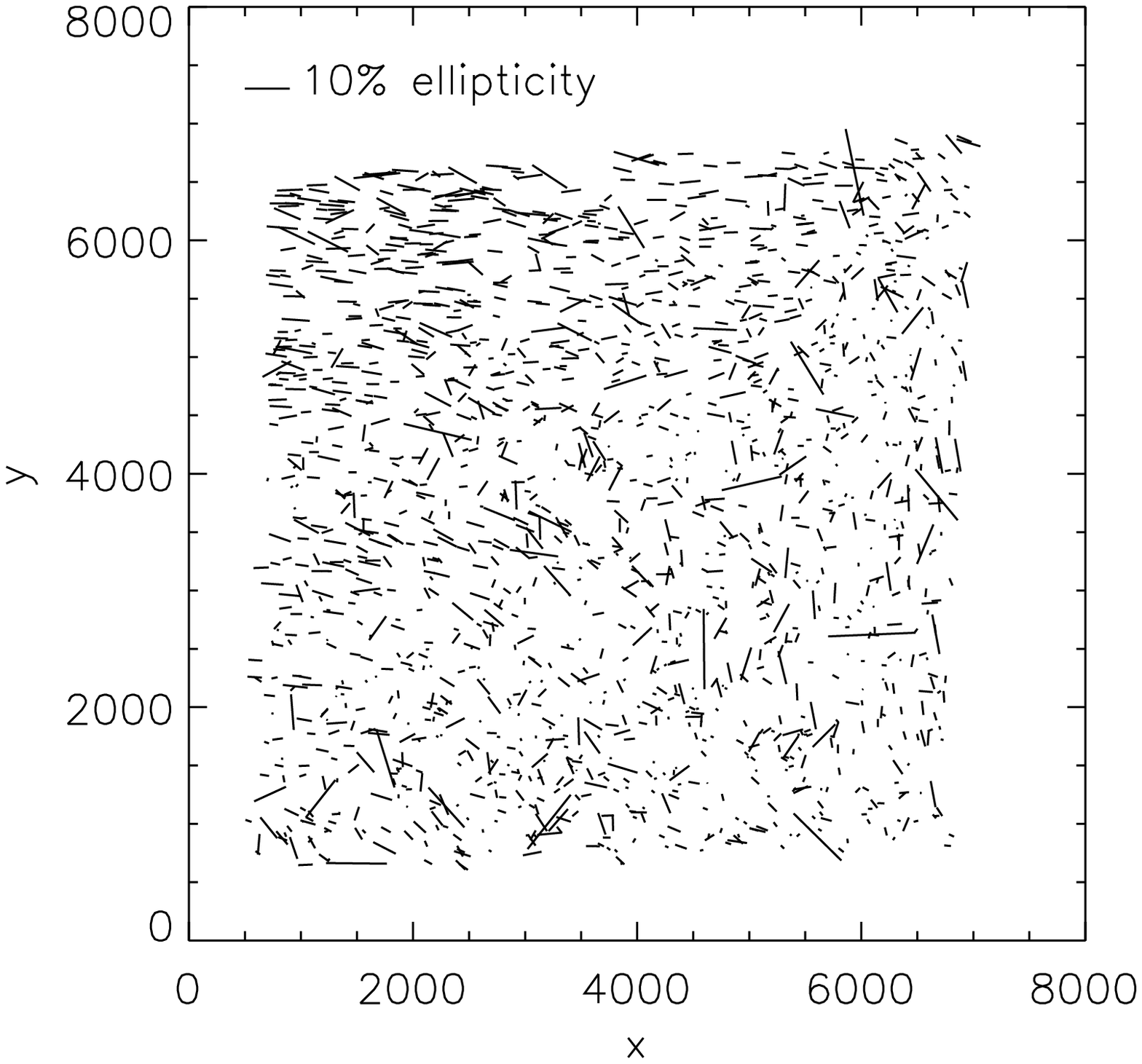}{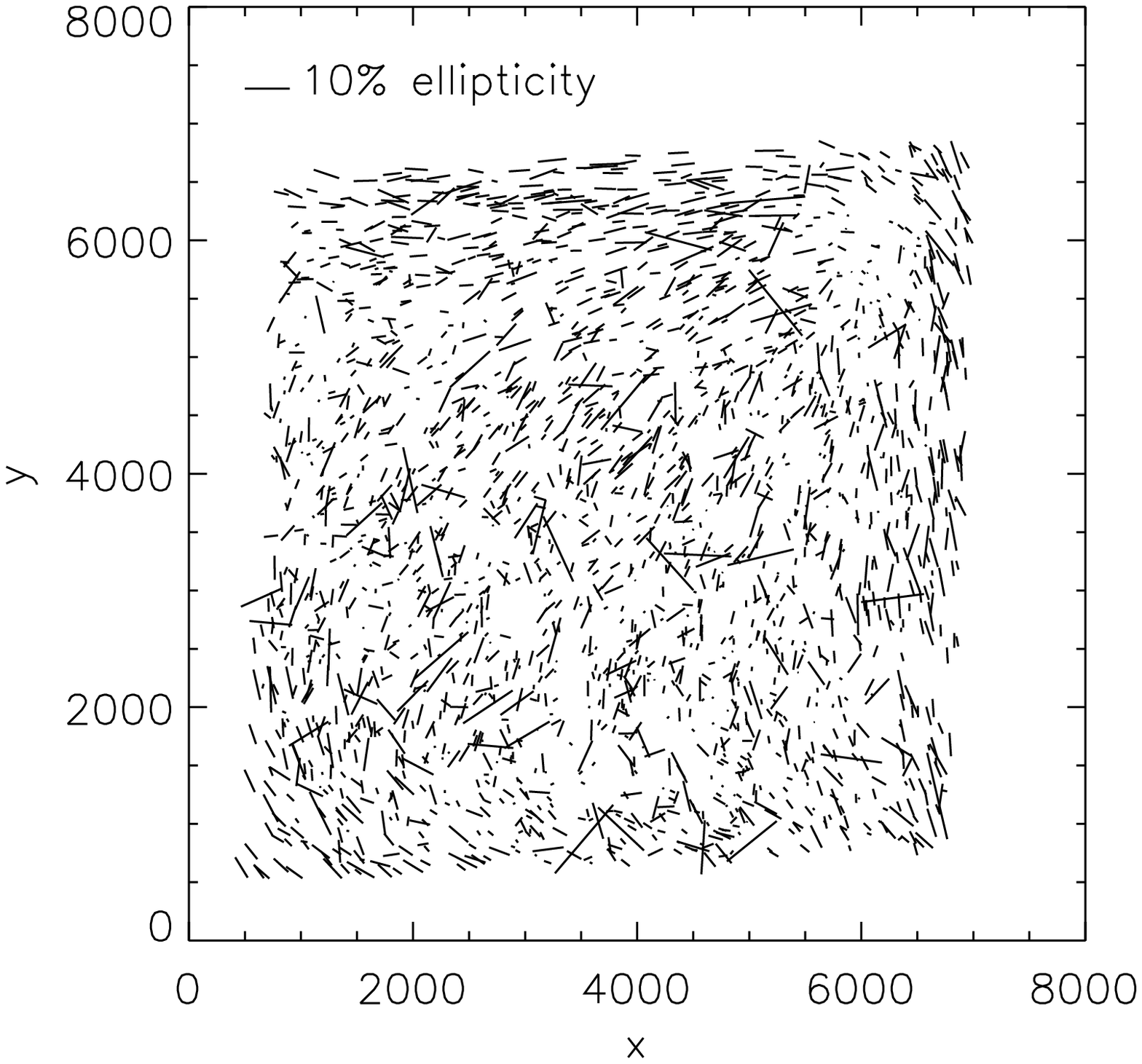}
\epsscale{1}
\caption{ These figures demonstrate that the  ACS PSF is time variable.  Each tick mark represents the magnitude and direction of the ellipticity of a star in the COSMOS field.  The left panel shows stars for the HST near the ``nominal'' focus value and the right panel shows stars in images taken when the primary-secondary mirror separation was 3 microns closer than nominal. Since the focus varies with time, the PSF is not temporally stable. While the plots are noisy, the different underlying patterns can be clearly discerned. }
\label{fig:psf_patterns}
\end{figure}
\begin{figure}[t]
\vspace{5mm}
\plotone{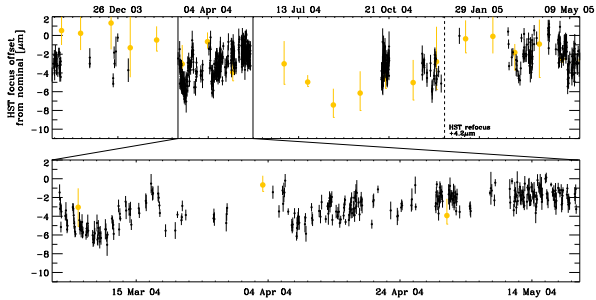}
\caption{The top panel shows the focus of HST for each field in the COSMOS survey(small data points). The errors bars are the error on the mean focus determined from the $\sim10$ useful stars in each COSMOS image.  The larger, fainter data points are focus values determined from a monitoring program using the ACS HRC (Lallo \etal\ 2006). Their error bars represent the full focus variation during an HST orbit.  The date of the 2004 secondary mirror movement is labelled.  The bottom panel shows the focus for a small time window during Cycle 12 (March to May 2004).  The cyclical pattern of the focus is clear in the bottom panel.  We describe how we determine the telescope focus value in \S\ref{models}}
\label{fig:focus}
\end{figure}
\begin{figure}[h]
\plotone{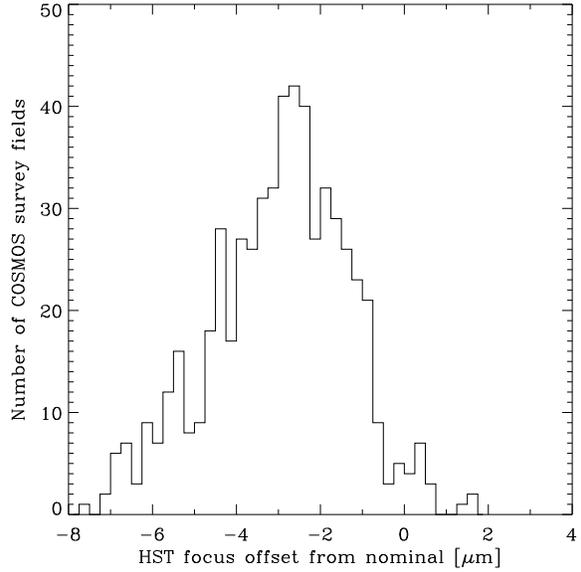}
\caption{A histogram of the best-fit focus values for each of the 575 COSMOS fields.}
\label{fig:focushist}
\end{figure}

The ACS WFC's PSF varies due to focus changes caused by thermal fluctuations of
the HST on orbit (Heymans \etal\ 2005; Rhodes \etal\ 2006; Schrabback \etal\
2006; Lallo \etal\ 2005; Lallo \etal\ 2006; Makidon \etal\ 2006a; Makidon \etal\
2006b).    This manifests itself in PSF patterns that vary significantly  over
time (Figure~\ref{fig:psf_patterns}).  Since the changes in focus are
stochastic, the focus of a particular observation cannot be predicted \emph{a
priori }and must be inferred from the data.  We have developed a method for
determining the telescope focus (see \S\ref{models}) and we show in
Figure~\ref{fig:focus} the focus of HST as a function of time over the course of
the COSMOS observations.  There are two time scales over which the HST focus
changes.  The first time scale is the 90 minute HST orbit; the telescope expands
and contracts, or ``breathes,'' as the telescope goes into and out of sunlight
(see, for instance, Lallo \etal 2006).  Since each COSMOS observation takes
place over a single orbit, we average over this source of focus variation in
COSMOS (and indeed in most HST applications).  This intra-orbit breathing is a
large source of the individual error bars on the points in
Figure~\ref{fig:focus} because the intra-orbit focus changes are of order a few
microns or more (Lallo \etal\ 2006). There is also an apparent slow oscillatory
drift in the focus of the telescope over several weeks; it is this gradual drift
that we attempt to model and correct for as described below.  From
Figure~\ref{fig:focus} it is clear that sampling the focus every month as done
by Lallo \etal\ provides insufficient sampling to satisfactorily model this 
drift.  We find that the focus is typically within $\pm2 \mu$m of $-3\mu$m, as
shown in Figure~\ref{fig:focushist}.  During the first decade of HST operations
it was necessary to periodically (approximately every 6 months) reset the
telescope focus as the struts holding the secondary mirror in place slowly
outgassed water and shrank.  However,  the last commanded focus changes of HST
were in December 2002, December 2004, and July 2006  because this slow focus
drift due to the shrinking of the structure has stabilized.

In order to examine the behavior of the ACS PSF we make use of two other HST data sets.  The first are  images of the stellar field 47Tuc taken as part of the ACS calibration  program (GO-10048; Krist 2003).  The three 47Tuc images are single 30 second exposures of dense stellar fields taken between 28 November 2003  and 7 September 2004 with the F814W filter.  These images are processed with the same pipeline we describe in \S~\ref{reduction} and we use these images to test our PSF models and focus prediction methods as described in \S~\ref{models}.

We also use Extended Pixel Edge Response (EPER) data from HST calibration
program 10369, consisting of images acquired using internal targets
(lamps) only, taken during Earth occultation time  (Mutchler \& Sirianni 2005).  These images overscan
the ACS CCD fiducial range of 2 $\times$ 2048 $\times$ 4096 pixels, providing additional
`virtual pixels' along the edges of the chips, into which pixel charge (but not actual
signal from the lamps) can transfer  during readout. The signal that occurs in the overscan region is charge that is released from charge traps during the readout process. This allows
calibration of the CTE of the ACS CCDs.  The EPER
images are 1.2 second exposures, taken using a combination of the F555W
and F435W filters to reduce the incident power from the lamps.  Note that the CTE occurs within the CCDs so is completely independent of the filters and only depends on flux, date, and CCD position (see \S \ref{cte}).

\subsection{Multidrizzle}\label{multi}
\begin{figure}[t]
\epsscale{1.5}
\plotone{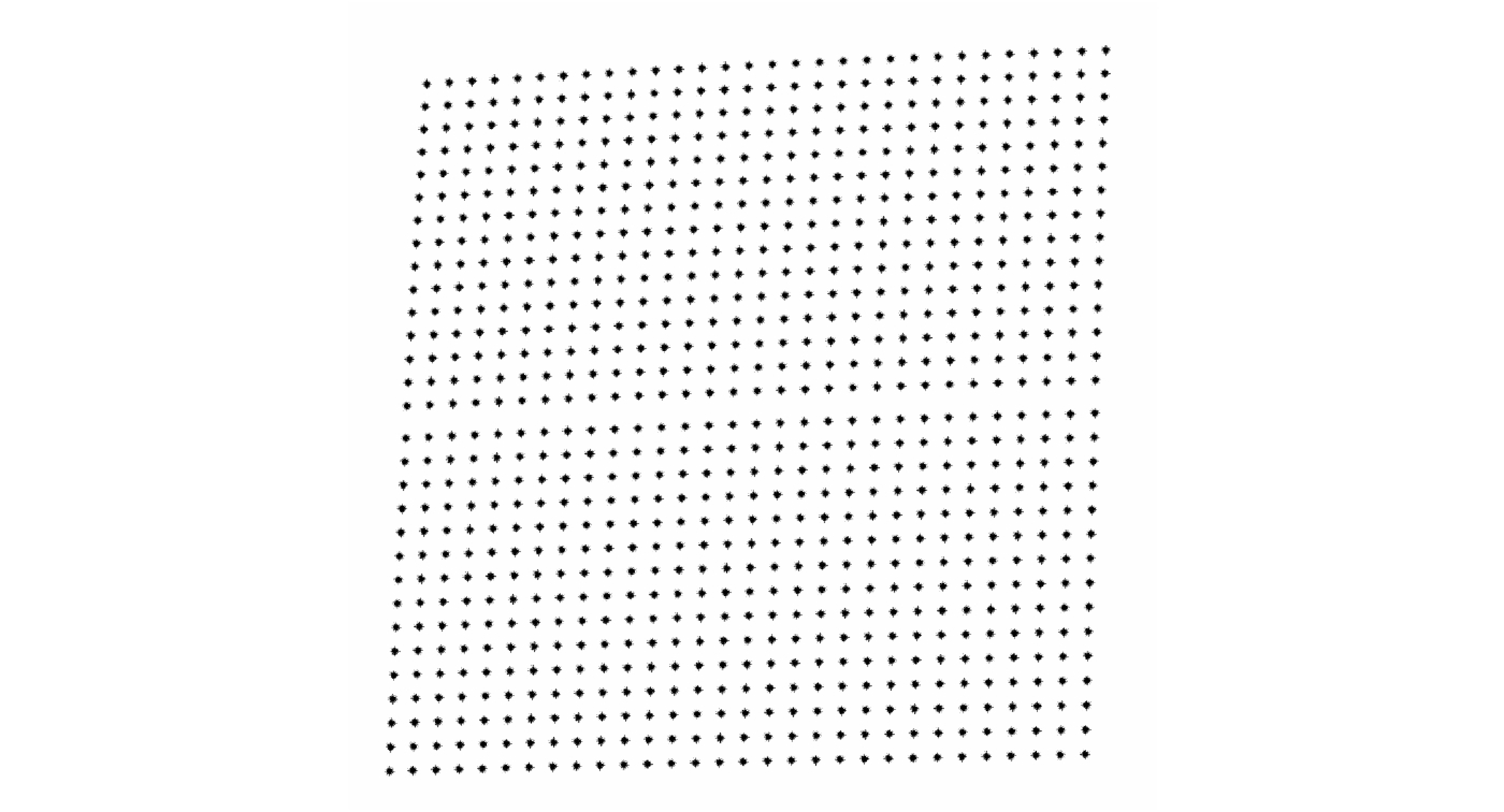}
\epsscale{1}
\caption{A typical artificial starfield created using our modified version of \tinytim. This field is at focus of $-2\mu$m.}
\label{fig:tinytim}
\end{figure}

The ACS is situated off-axis in the HST focal plane and the ACS focal plane is not normal to incident light rays.  Therefore ACS images have a large geometric distortion (Anderson 2006; Pavlovsky \etal 2006).  The $4096\times4096$ ACS WFC CCDs do not sample a square on the sky, they sample a ``squashed parallelogram'' (see, for example, Figure~\ref{fig:tinytim}.  The program \multidrizzle\ removes this geometric distortion while simultaneously removing cosmic rays and bad pixels as well as combining multiple exposures into a single output image (Koekemoer \etal\ 2002).  
The geometric distortion  is easily measured and corrected for with astrometric observations of stellar fields, as shown by Meurer \etal\ (2002) and Anderson  (2006).  They fit the  geometric distortion  to a quartic polynomial and the resulting fit is good to better than 0.05 pixels across the ACS WFC.  Subsequent observations have improved this fit and the latest distortion polynomial coefficients are readily available from the Space Telescope Science Institute.  Despite the fact that the geometric distortion  is the largest PSF effect, it is the easiest to correct for and we use the default correction available via \multidrizzle.  The geometric distortion is not completely temporally stable and, indeed, has changed over the lifetime of the ACS (Anderson 2006).  However, the size of this variation is much smaller than the variation in PSF due to changes in the telescope's focus.  The latest geometric distortion corrections are used in the reduction of the COSMOS data (Koekemoer \etal 2006).
The pixel scale of the \multidrizzle\ output image can be smaller than the pixel scale of the input image (or images).  The details of how \multidrizzle\ is run on COSMOS data are given by Koekemoer et al. (2006).  We make some important changes in the default \multidrizzle\ parameters in order to optimize the images for weak lensing analysis as described in \S~\ref{reduction}.


\subsection{TinyTim}\label{tt}

There are not enough stars in each extragalactic COSMOS ACS image to allow us to model the PSF across the field through interpolation.  Therefore, we have simulated artificial stars in ACS fields at arbitrary positions (in the following discussion we will use the term `star'  and `PSF' interchangeably) using  the \tinytim\ software package (Krist \& Hook
2004).  \tinytim\  can create  PSFs for any current HST camera with any  filter combination, at any detector position, and for any given input spectrum.  \tinytim\ creates FITS images
containing  stars that include the effects of diffraction, geometric
distortion, and charge diffusion within the CCDs.    By default, the stars appear
as they would in raw  HST images.  In the case of ACS, this means the stars are highly distorted and have a pixel scale
of 0.05 arcseconds per pixel. \tinytim\ is able to make highly oversampled PSFs and can incorporate the focus position (primary/secondary mirror spacing of the telescope).

We have adapted version 6.3 of the \tinytim\ software package  to create simulated ACS  starfields.  By default, ACS creates only single stars or several stars in a small portion of the ACS WFC.  Using the \texttt{IDL}  programming language, we have written a wrapper  that allows us to run \tinytim\ multiple times and create a grid of
PSF models across the whole ACS field of view.
This is available from   \\ {\url http://www.astro.caltech.edu/$\sim$rjm/acs}. We insert our artificial stars
into blank images with the same dimensions and FITS structure as real ACS data,
thereby manufacturing arbitrarily dense starfields. The default \tinytim\ pipeline calculates a diffraction pattern (spot diagram), geometrically distorts this pattern, and adds charge diffusion between adjacent pixels.    The geometric distortion coefficients built into \tinytim\ were current at the time of the program's release but have since been superseded.  Our \texttt{IDL} \tinytim\ wrapper allows us to input the most recent distortion coefficients.   These three  effects occur (in that order) in real data and  depend upon the
position of the star in the ACS field of view. \tinytim\ is run in steps, and, with our  modifications, the \tinytim\ software allows some of these steps to be skipped.
By skipping various steps within our \texttt{IDL} wrapper we  allow for two important changes to this basic pipeline. The
deviations from the default \tinytim\ pipeline are:

\begin{itemize}

\item In order to examine the effects of the distortion removal process (\multidrizzle\ in our case), we incorporate the option to allow each star to have an identical
diffraction pattern and charge diffusion, but a geometric distortion determined
by the location of the PSF within the ACS field of view. Once the geometric
distortion is removed by running the field through \multidrizzle\,
these stars should all appear identical.

\item We allow \tinytim\ to create  starfields that do not contain the effects of geometric distortion at
all,  instead modelling stars as they would appear after a perfect removal of
geometric distortion. Conversion between
distorted and non-distorted frames, which is necessary to simulate charge
diffusion in the raw CCD, is performed using very highly oversampled images.
This avoids stochastic aliasing of the PSF (see \S\ref{reduction}), and
minimizes noise in the PSF models. This makes use of the geometric distortion coefficients described in Gonzaga et al. (2005).

\end{itemize}

Our modified version of  \tinytim\ thus allows us to make dense artificial
starfields at a range of focus values, and is publicly available from \\ {\url
http://www.astro.caltech.edu/$\sim$rjm/acs}.  We have created starfields
in the F814W filter for the COSMOS analysis, but the procedures we have
developed are general and can be adapted to any filter. For reasons of speed, we
raytrace through the optical path at a single wavelength of 814nm. Using a full
stellar (or galaxy) spectrum in the F814W filter does not significantly change
the output PSF but adds considerably to the processing time. We generate the PSF models in an oversampled pixel frame. This can be repixelated to the ACS pixel size in geometrically distorted (raw) coordinates or in geometrically corrected coordinates. In the distorted coordinates, we find that the simulated PSF is slightly smaller than real stars, possibly because of pixelization effects. This is fixed by convolving the PSF model with a square kernel the same size as the pixels.  Note that, as discussed in \S\ref{data}, our \tinytim PSF models do not include all PSF effects, especially near the red end of the spectrum.

We also have included the ability to create a set of raw images with an arbitrary
dither pattern.  Stars can be added to each geometrically distorted dithered
image in such a way that all the stars line up with each other on the output
image.  This portion of our \tinytim\ pipeline makes use of the
``\texttt{wtranback}'' coordinate transformation routine built into
\texttt{pyraf}. We can then make four dithered images with, for instance, the
COSMOS dither pattern and then combine them into a single output file using
\multidrizzle. See, for example, Figure~\ref{fig:aliasing1}. This is non-trivial because the large geometric distortion in ACS
can substantially change the relative  spacing of stars in dithered images.  We
describe the results of such tests in the following section.

\subsection{Shape Measurement}\label{rrg}

Here we introduce our formalism for the measurement of galaxy shapes. The method of Rhodes, Refregier, \& Groth (2000; hereafter RRG) has been optimized for space-based images with  small PSFs and has previously  been used on weak lensing analyses of WFPC2 and STIS data (Rhodes, Refregier, \& Groth 2001; Refregier, Rhodes, \& Groth 2002; Rhodes \etal\   2004).  Following the formalism of RRG, we parameterize object shapes by measuring the Gaussian weighted second order moments:

\begin{equation}
\label{eq:moments}
I_{ij}=\frac{\sum w I x_i x_j }{\sum w I}.
\end{equation}

\noindent The summation is over all pixels, $w$ is the size of the Gaussian weight function,  $I$ is the pixel intensity, and the coordinates $x_i$ are measured in pixels.  These moments are used to derive the ellipticity of an object $e_i$ and a size measure $d$ given by

\begin{equation}
\label{eq:e1}
e_{1}=\frac{I_{xx}-I_{yy} }{I_{xx}+I_{yy}},
\end{equation}

\begin{equation}
\label{eq:e2}
e_2=\frac{2I_{xy}}{I_{xx}+I_{yy}},
\end{equation}
and
\begin{equation}
d=\sqrt{\frac{(I_{xx}+I_{yy})}{2}}.
\end{equation}

\noindent The RRG method also requires the measurement of the five fourth order moments given by:

\begin{equation}
\label{eq:moments4}
I_{ijkl}=\frac{\sum w I x_i x_j x_k x_l}{\sum w I}
\end{equation}

\noindent in order to correct for the effects of the weighting function and the shape distortions caused by the PSF.  These fourth order moments are used to make small corrections to the second order moments.  In this way, the RRG method is a bridge between the earlier KSB (Kaiser, Squires and Broadhurst 1995) method which uses only second order moments, and more advanced methods such as Shapelets  (Massey \& Refregier 2005) which uses moments to arbitrarily high order based on the amount of information available in each object.

The ellipticities are related to the shear, the quantity of interest for weak lensing, via the shear susceptibility factor $G$:
\begin{equation}
\label{eq:shear_susc}
\gamma_{i}=\frac{e_i}{G}.
\end{equation}
Previous weak lensing analyses using the RRG method have made use of a single value of $G$ for the entire survey. We find that residual PSF systematics are reduced when $G$ is allowed to vary as a function of object size $d$, ellipticity $e$ and signal-to-noise.
Leauthaud et al. (2006) contains a discussion of the calculation of $G$ for the COSMOS data set.  Given an output pixel size of 0.03 arcseconds (see \S\ref{reduction} and  the \texttt{Source Extractor} parameters in Leauthaud \etal (2006), we find that the optimal weight function size for COSMOS data is
\begin{equation}
\label{eq:w}
w=\textrm{maximum}[2\sqrt{ab},6]
\label{eq:weight}
\end{equation}
\noindent where $a$ and $b$ are the \texttt{Source Extractor} computed semi-major and semi-minor axes, and the minimum weight function width 6 has been empirically determined to be the optimal weight function to measure the shapes of stars.
These above parameters (especially the factor of 2 in Equation~\ref{eq:weight}) have been tuned empirically and depend on the \texttt{Source Extractor} settings.

\section{Optimal Image Reduction}\label{reduction}
The transformation of pixels from a distorted input image to an undistorted output plane can introduce significant ``aliasing''
of pixels if the output pixel scale is comparable in size to the input scale. When
transforming a single input image to the output plane, point sources can be
enlarged, and their ellipticities changed significantly depending upon
their sub-pixel position. This is one of the fundamental reasons why dithering
is recommended for observations,
since the source is at a different sub-pixel position in different exposures,
thus partially mitigating these effects  when exposures are
combined. However, the residual aliasing  in combined images is
sufficient to prevent the measurement of small, faint galaxies at the
precision required for weak lensing analysis.

\begin{figure}[h!]
\epsscale{0.48}
\plotone{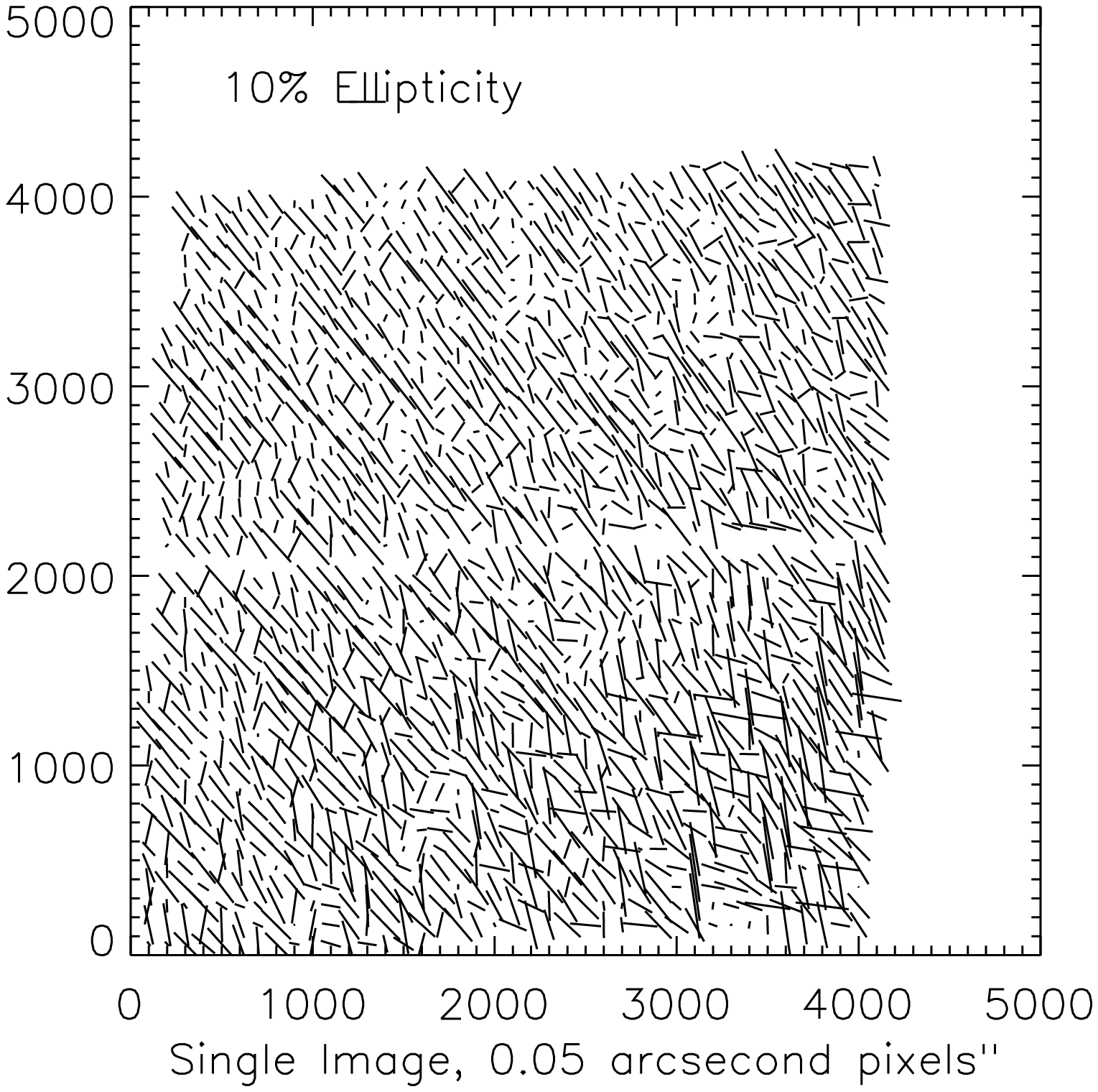}
\plotone{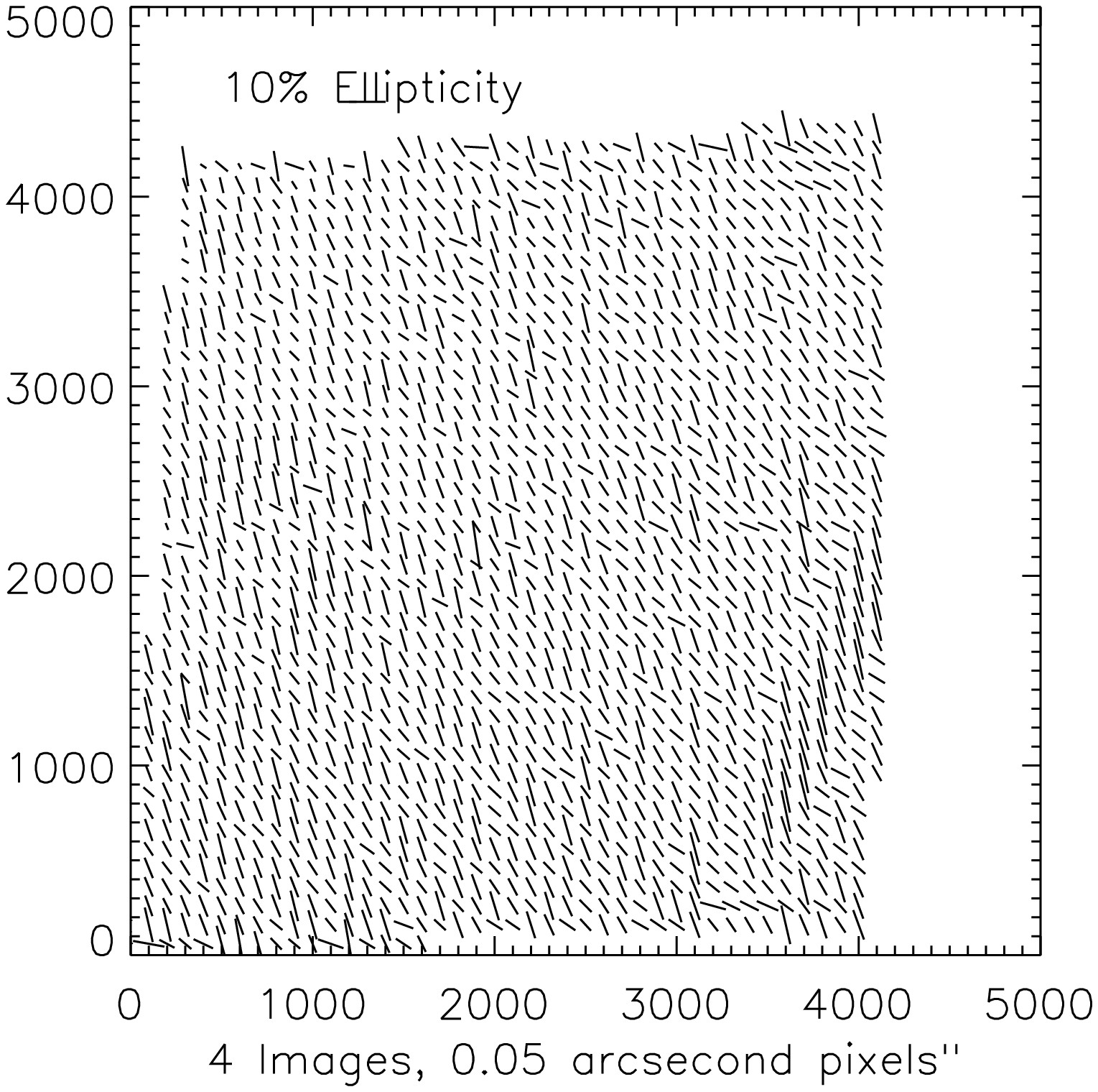}
\plotone{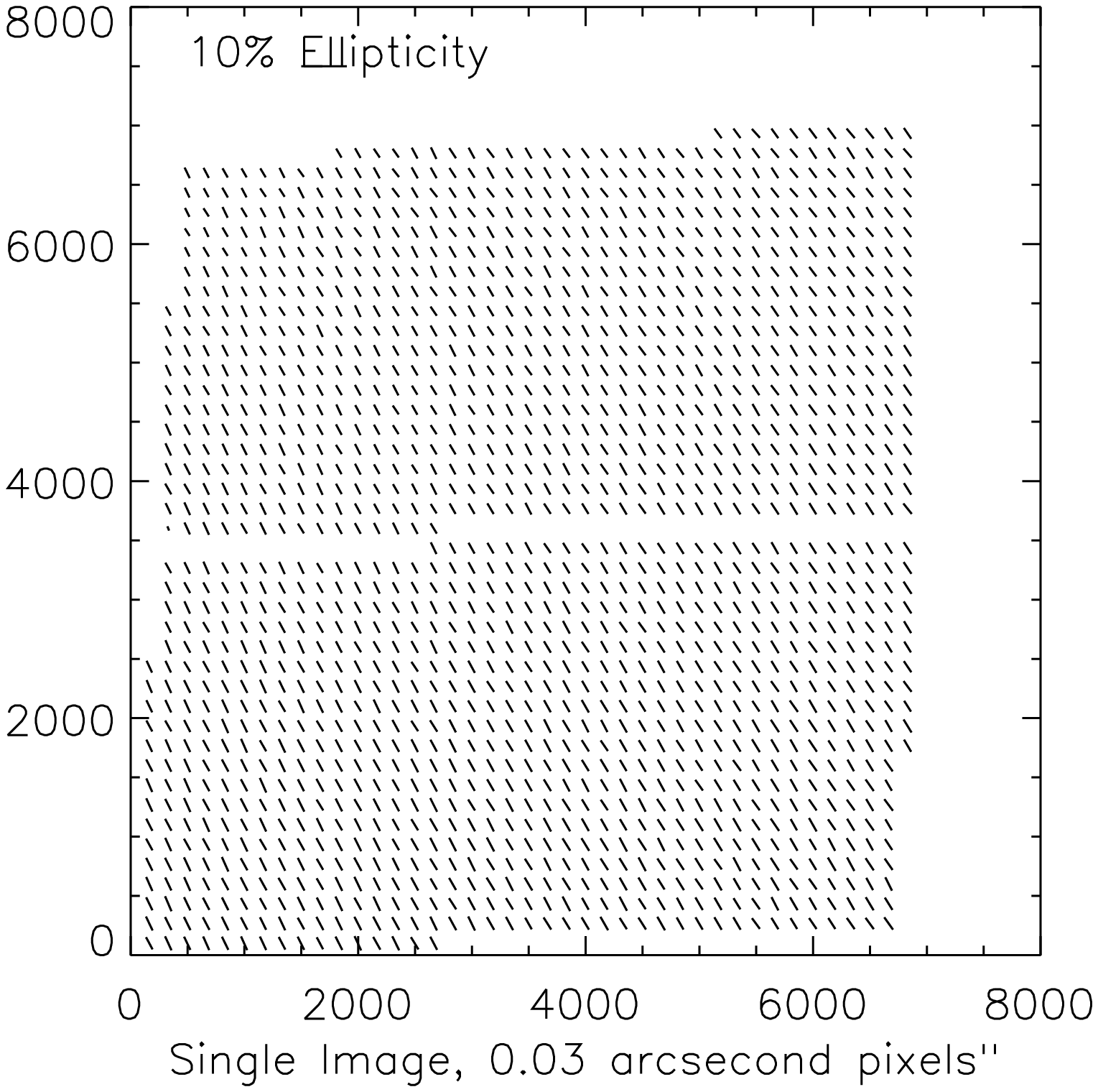}
\epsscale{1.0}
\caption{The ``aliasing'' of the PSF introduced when transforming  distorted
input images to an  undistorted output frame. Each tick represent the ellipticity of  a star that has undergone
identical diffraction in
the telescope's optics.  Each tick mark should therefore look  identical. The difference
between stars is the  sub-pixel position when  geometric distortion is
removed via the  \multidrizzle\ program.  The first panel shows a single image run through \multidrizzle\ with the default settings.   This problem can be ameliorated by altering
several of the \multidrizzle\   settings and using   dithered input images.  The second panel shows that the ellipticity scatter is reduced when four dithered images are combined with the default \multidrizzle\ settings.  We show in Figure~\ref{fig:kernels} and in the third panel that reducing the output pixel
scale greatly reduces this source of PSF uncertainty.  This panel shows the results when a single input image is processed using our optimized \multidrizzle\ settings, including a final output pixel scale of $0.03$ arcesconds. The scatter in the output ellipticities is further reduced by using 4 dithered images with optimized \multidrizzle\ settings. We do not show this plot as the reduced ellipticity scatter is difficult to detect by eye.
The inter-chip gap visible in the first and third panels is covered by the dither pattern, which was one of the goals of the COSMOS dither pattern.}
\label{fig:aliasing1}
\end{figure}

In order to demonstrate the effects of aliasing we created a set of \tinytim\ starfields that contain the same diffraction and diffusion for each star but a geometric distortion given by the position of the star within the ACS WFC field.  We then use \multidrizzle\ (with  default settings) to remove this geometric distortion. The resulting, undistorted field should contain stars that are identical in all portions of the chip.  We show in the first panel of Figure~\ref{fig:aliasing1} that this is not the case.  In this  figure, each tick mark represents a \tinytim\ created star.  The length and orientation of the tick mark represents the size and direction of the star's ellipticity as given by Equations~\ref{eq:e1} and \ref{eq:e2}.  The fact that the tick marks vary is evidence of aliasing.  We have found that dithering and reducing the size of the output pixels  reduces the scatter in the ellipticities of stars in this study.  The largest gain comes from reducing the size of the output pixels, but, as expected, the scatter in $e$ also drops as roughly $\sqrt{N}$ for $N$ dithers. This
confirms the idea that the repixelization adds stochastic noise to the
ellipticity when the  sub-pixel positions are uncorrelated.
 Thus, sub pixel aliasing is not a problem when there are many dithers (e.g. the Hubble Ultra Deep Field) but for fields with a limited number of dithers like COSMOS, this is a significant source of PSF error.  As we show below, this effect can be minimized by  carefully choosing several \multidrizzle\ parameters.

\begin{figure*}[p]
\epsscale{1.3}
\plotone{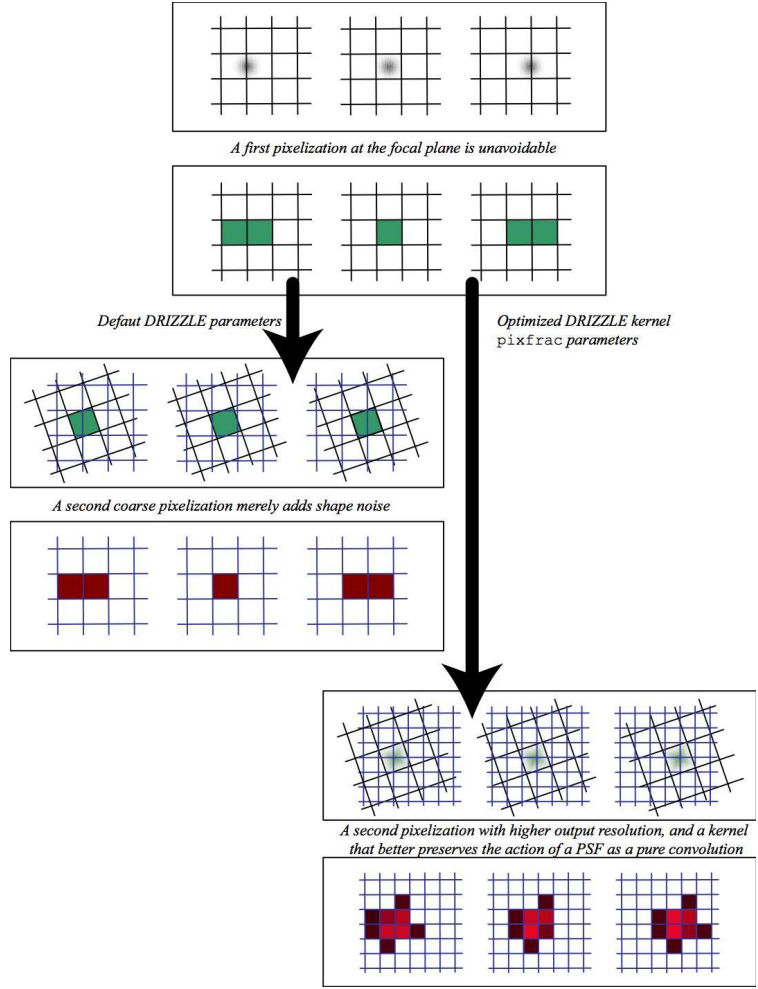}
\epsscale{1.0}
\caption{{ This series of illustrations depicts the aliasing  of images caused by
pixelization. The first aliasing effect (depicted in the top two rows) is an
inevitable consequence of observing the continuous sky with discrete pixels. The
top row of this panel shows a simulated object on a pixel grid, and the second
row shows how that object would appear in a pixelized image. Point sources that
lie near the center of a pixel are mostly detected in that pixel, as shown in
the center column.  However, if the center of an object lies near the border
between two pixels, the object is elongated in one direction during the process
of observation.  Such an elongation mimics the shear caused by weak lensing and
is shown in the left and right columns of the first two rows. The middle tow
rows illustrate a second pixelization that can change the shapes of observed
objects.  This pixelization occurs during the  \multidrizzle\ stage of image
processing when images are resampled onto a new output grid in order to remove
geometric distortion.  In each of the three instances here, a perfectly centered
star in the original (black tilted) coordinate frame is drizzled onto the
underlying output pixel grid, which is aligned with the page.  A second
aliasing  occurs at this stage if the input pixel lies on the border of two
output pixels.  Again, the middle columnn shows no aliasing but the left and
right images show the object being elongated in one direction.  The image
distortion in the second pixelization is, however, somewhat avoidable.  If the
output pixel grid is made smaller, as shown in the bottom two rows, the shape of
the object on the final pixel grid is much less dependent on the relative
alignments of the input and output pixels. Sub-pixel dithering of  many input
images and subsequent image combination onto a fine output pixel grid can
further reduce the effects of pixelization.}}
\label{fig:aliasing2}
\end{figure*}

Pixelization effects are unavoidable during the initial
exposure, when  the detector discretely samples an image. However, it is clearly
desirable to minimize related effects during data reduction. The effect on each
individual object depends on how the input and output pixel grids line up. These arguments are demonstrated graphically in Figure~\ref{fig:aliasing2}.
That figure demonstrates that we can  mitigate this effect  by using a finer grid of output pixels ({\it
e.g.} Lombardi et al. 2005). This reduction in pixel scale (which will cause a
corresponding increase in computer overheads) can be performed in conjunction
with simultaneously ``shrinking'' the area of the input pixels that contains
the signal, by making use of the \multidrizzle\ \texttt{pixfrac} parameter
and convolution kernel.

\begin{figure}[tb]
\epsscale{.9}
\plotone{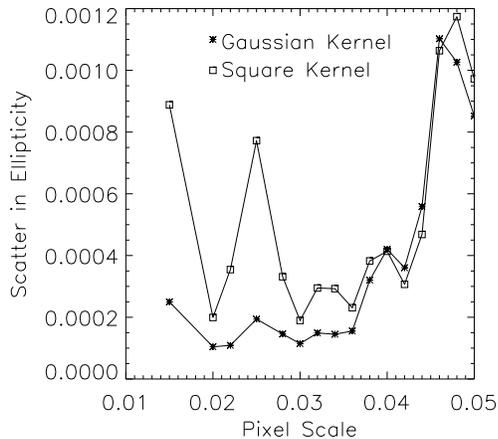}
\epsscale{1}
\caption{RMS ellipticity introduced during \multidrizzle.
 Lower values of the scatter in ellipticity show more stable behavior of the PSF during this stage of image combination and geometric distortion removal. The only sources of $e$ variation here is pixelization and measurement errors; each star was designed to have the same PSF after the removal of geometric distortion.  Each measurement represents a grid of $20\times20$ stars spread across the ACS WFC field.  The scatter in the ellipticity is measured as the standard deviation of the mean of measured ellipticity values.  Note the resonance in the square kernel at 0.025 arcsecond pixels, half the original pixel size.  The reduction of output pixel scale is the most important improvement we make on the default \multidrizzle\ pipeline, with a slight further gain from going to a Gaussian rather than square  kernel.
 For the COSMOS weak lensing images, we  use a Gaussian kernel and an output pixel
size of 0.03 arcseconds (Leauthaud \etal 2006).  This  minimizes the effects of undersampling on
the PSF and produce images that are optimal for weak lensing science at the cost of
 introducing significant additional correlated noise
relative to the square kernel. This  is not critical for weak lensing science
but may not be optimal for other types of science.}
\label{fig:kernels}
\end{figure}

We have run a series of tests on the simulated PSF grids
 to determine the optimal values of the \multidrizzle\
parameters specifically for weak lensing science. As described above, we  produced a grid of stars that ought to look
identical after the removal of geometric distortion.
We then ran a series of tests using \multidrizzle\ on the same input image
but with a range of output pixel scales, convolution kernels, and values of
\texttt{pixfrac}. We  measured the scatter in the ellipticity values in the
output images. The smaller the scatter, the more accurately the PSF is
represented. We found the results were not strongly dependent on the choice
of \texttt{pixfrac} and settled on \texttt{pixfrac}$=0.8$.  We show in
Figure~\ref{fig:kernels} that PSF stability is improved dramatically by reducing
the output pixel scale from 0.05 arcseconds (the default) to 0.03 arcseconds.
There is  a very slight gain in going to smaller output pixel sizes but the
storage requirements increase rapidly as the number of output pixels is increased. The gain in going to smaller
pixel scales is more stable with a Gaussian kernel  than
with the default square kernel. Therefore, for weak lensing  we use  an output pixel scale of
0.03 arcseconds, \texttt{pixfrac}=0.8, and a Gaussian kernel. Despite its
clear advantages for weak lensing studies, the Gaussian kernel does have some
general drawbacks, such as the introduction of more correlated
noise which may not be desirable for other types of science where minimization
of correlated noise is important.

\section{Focus Dependent PSF models}\label{models}

Changes in the HST's focus  significantly change the PSF. These focus changes are caused by a change in the primary-secondary mirror spacing brought about by thermal fluctuations.  This spacing (which we hereafter refer to as the focus) can deviate from its nominal value  in the range $-10\mu m$ to $+5 \mu m$.  We thus want to know what the PSF looks like at each point in the ACS WFC field for this entire range of focus values.  The repixelization  of ACS data necessary  to remove geometric distortion  causes stochastic aliasing of the PSF even with the  optimal \multidrizzle\ parameters presented in \S~\ref{reduction}.  Because of this, we create \tinytim\ stellar fields without geometric distortion. These contain stars that have diffraction and diffusion given by their position within the ACS WFC field, but appear as they would after a perfect removal of geometric distortion. These are stars as they would appear if the number of exposures $N$ was very large.  Since we are creating simulated stars, we make the spacing between stars small enough that we do not have to do a complicated interpolation between stars to find the PSF at a particular point in the ACS WFC field.  We simply choose the closest star. We find that this condition is satisfied if we create grids of 30$\times$ 30   stars across the two ACS WFC CCDs.  We also find that creating such a grid in the focus  range $-10$ to $+5\mu$m at one micron increments
gives sufficient resolution in focus that the difference in ellipticity between a star at two adjacent focus values (at the same chip position) is smaller than the measurement error in measuring the ellipticity of a star in a typical noisy COSMOS image.

\begin{figure}[h]
\plottwo{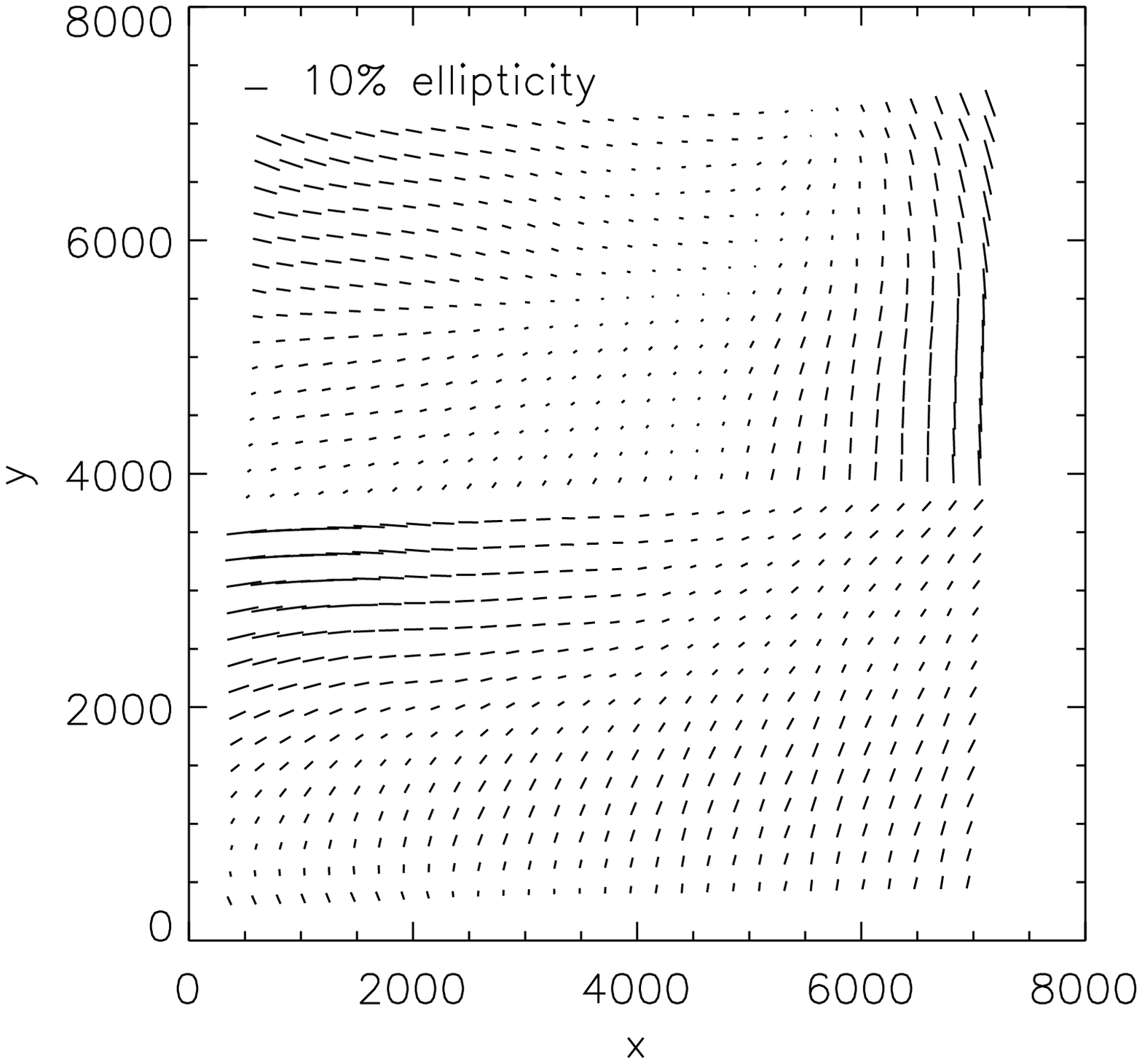}{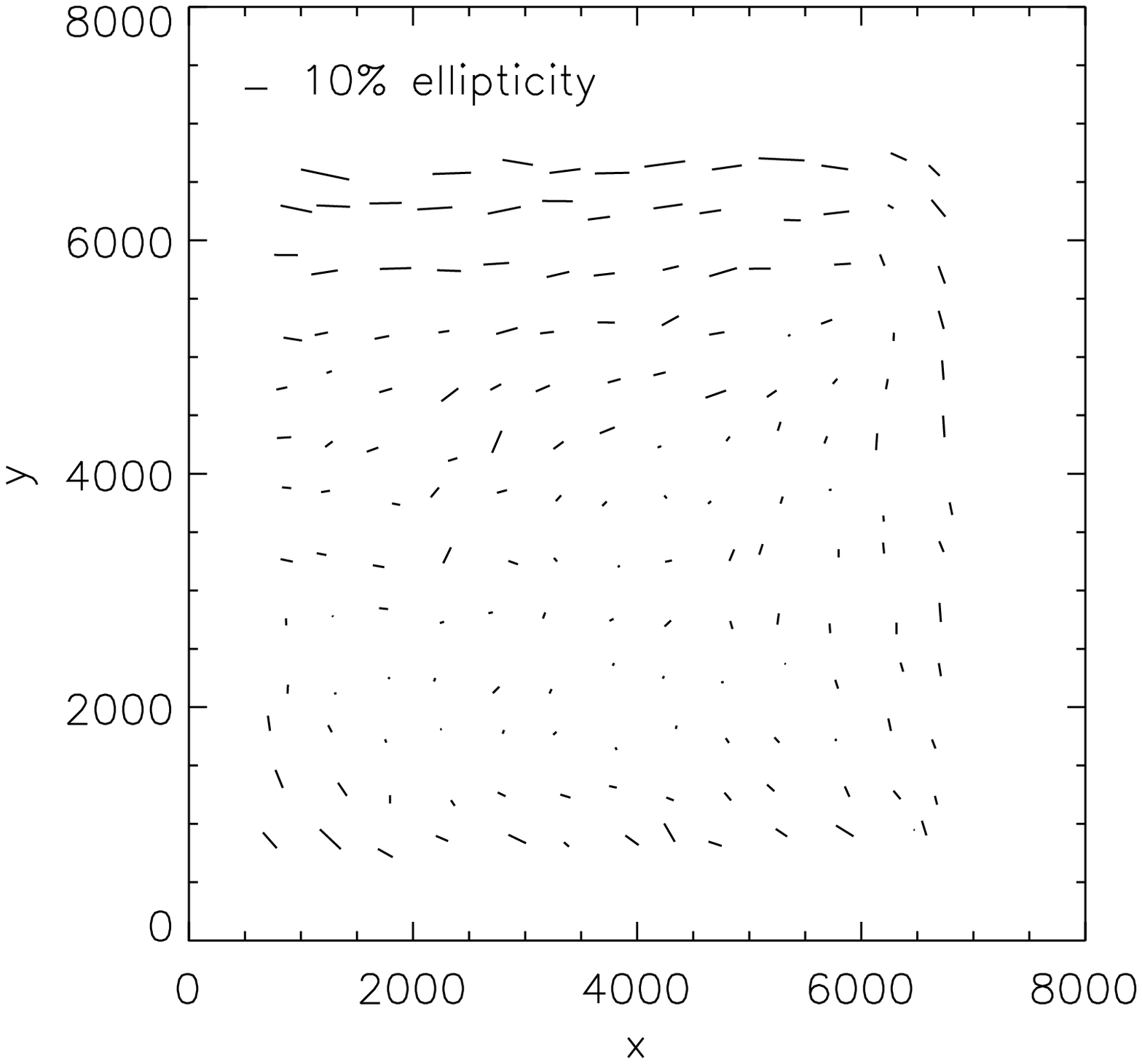}
\caption{ A  \tinytim\ PSF model (left panel) for a focus value of $-2\mu$m and the average of many observed stars (right panel) from COSMOS fields with a similar apparent focus.  There is rough qualitative  agreement between the data and the models over much of the ACS field. The  center of the chip does not show good agreement due, at least in part, to the effects of degradation of the CTE. The real data show less positive $e_1$ (elongation along the $x$ axis) than the models, consistent with a CTE-induced smearing in the $y$ direction.  We remove this instrumental signature late in our lensing analysis as described in \S\ref{cte}  Notice that the $x$ and $y$ ranges are not $[0,4096]$ as in default ACS images but are larger because we are using a smaller output pixel scale.  These plots are not meant to demonstrative quantitative agreement between the models and the real stars. The \tinytim\ models are noiseless, while the real COSMOS stars have image noise and each 'tick mark' represents a different number of stars (usually about 10 or fewer). For the RRG method, it is important for the \tinytim\ stars to match the moments (rather than the more easily plotted ellipticities) of the real stars.  A quantitative analysis of how well the model stars can be used  to deconvolve the PSF can be obtained from the star-galaxy correlation functions in Figure~\ref{fig:stargal} and in Massey \etal 2006a.}
\label{fig:acs_tinytim}
\end{figure}

In a typical COSMOS field approximately 10 stars have a suitable S/N and are sufficiently deblended from other objects  to provide accurate PSF measurements.  See Leauthaud \etal\ 2006 for a description of star-galaxy separation in the COSMOS weak lensing catalog.  We compare these stars to each focus model in the range $[-10,5]\mu$m.
We determine the \tinytim\ focal position that minimizes the $\chi^2$
between the ellipticities of the stars in data and the corresponding
closest stars in the \tinytim\ model, where the $\chi^2$  is defined as
\begin{equation}
\label{eq:goodfit}
\chi^2=\sum(e_{1}^{*}-e_{1}^{TT})^2+(e_{2}^{*}-e_{2}^{TT})^2
\end{equation}
\noindent The superscripts * and \texttt{TT} represent real COSMOS stars and \tinytim\ simulated stars, respectively, and the sum is over all the stars in the image that make the cuts described above.  Thus, we determine the best fit for the telescope's focus at the time of the observation. This is the focus value we show in Figure~\ref{fig:focus}.  Note that this method of determining focus necessarily averages over any intra-orbit focus changes.  Figure~\ref{fig:acs_tinytim} shows the stellar ellipticity pattern in the ACS WFC for a value of focus=$-2\mu$m alongside the pattern formed by averaging all of the approximately 2000 COSMOS stars determined to be at that focus value.     There is fair qualititative agreement between the data and the model.  This agreement is poor  in the center of the ACS WFC field.  We have determined that part of this  disagreement is  due to CTE degradation, which is not included in the \tinytim\ PSF models.  We discuss the causes of this and our solution in \S~\ref{cte}. the shown \tinytim\ model represents the best match to the ACS COSMOS data from all the focus values for which we calculated \tinytim\ focus models.  The real test of the quality of our PSF models is in the star-galaxy correlation functions before and after PSF correction.  These are shown in Figure~\ref{fig:stargal}.


\begin{figure}[h]
\plotone{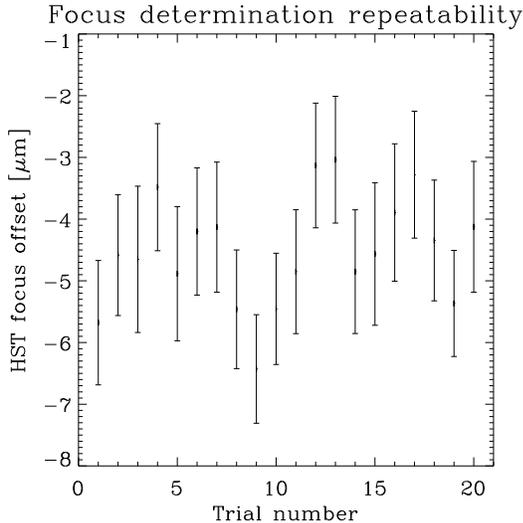}
\caption{Determination of the focal position from 20 sets of 20 independent stars,
each chosen randomly from a single image taken of a stellar field near the
globular cluster 47Tuc.  The x-axis shows the set number, from 1 to 20,
and the y-axis shows the determined focal position.  The uncertainty on
the y-axis is the measured uncertainty of the focal position.  As is
shown, the focal position determination is highly repeatable in random
trials of independent stars at identical focal positions. From these sorts of tests, we estimate that the uncertainty in determining the focal position for a short exposure is approximately $\pm 1\mu$m. The determination of the focus from a a set of exposures taken over a full orbit has a greater uncertainty due to the intra-orbit `breathing' of the telescope.}
\label{fig:47TUC}
\end{figure}

We have tested our focal position determination using randomly selected
stars taken from the calibration images of 47Tuc described in \S~\ref{data}.  We select 20
random non--overlapping sets of 20 stars from the 746 stars seen in one of the 47Tuc images.  We determine the focus for each of these sets of
stars.  As these stars are all from the same image, the focal position
determined should be identical for all sets.  We find that all of the determined focal positions are within $\pm 2 \mu$m of the mean
focal position. Based on these simulations, the results of which are shown in Figure~\ref{fig:47TUC}, we estimate the uncertainty with which we can measure the focal position in a short exposure to be about $\pm 1\mu$m.  The uncertainty in determining the focus value for a COSMOS image is somewhat larger than this because we average over the intra-orbit focus changes caused by telescope `breathing' (see Figure~\ref{fig:focus}).


\section{PSF Correction in COSMOS Images}\label{correction}

We follow the  procedure given in RRG (2000) to correct galaxy shapes for PSF.  Stellar moments are used to correct galaxy moments first for the isotropic portion of the PSF, then for the anisotropic portion.  Quantities are kept in terms of moments (rather than ellipticities)  through the entire correction process and only then are the corrected moments used to calculate galaxy ellipticities, and, ultimately, shear.
We tested using a variable stellar weight function width $w$ instead of the standard (for RRG) fixed stellar weight function width and found that this did not significantly improve the PSF correction.  The weight function used to evaluate galaxies still varies with the size of the galaxy as discussed in \S~\ref{rrg}.

\subsection{Applying the TinyTim models}\label{tinytim}

\begin{figure}[t]
\epsscale{1.05}
\plottwo{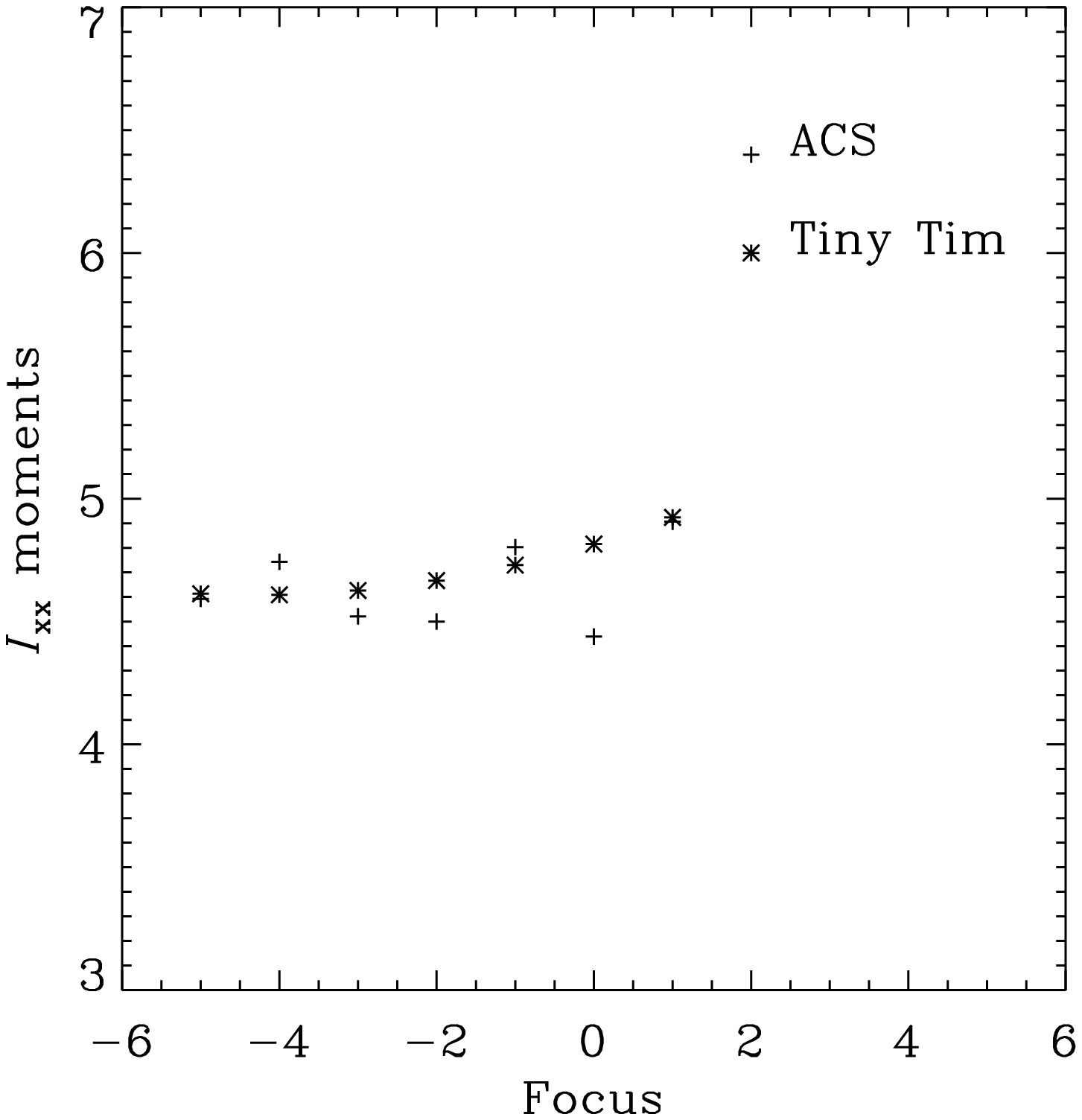}{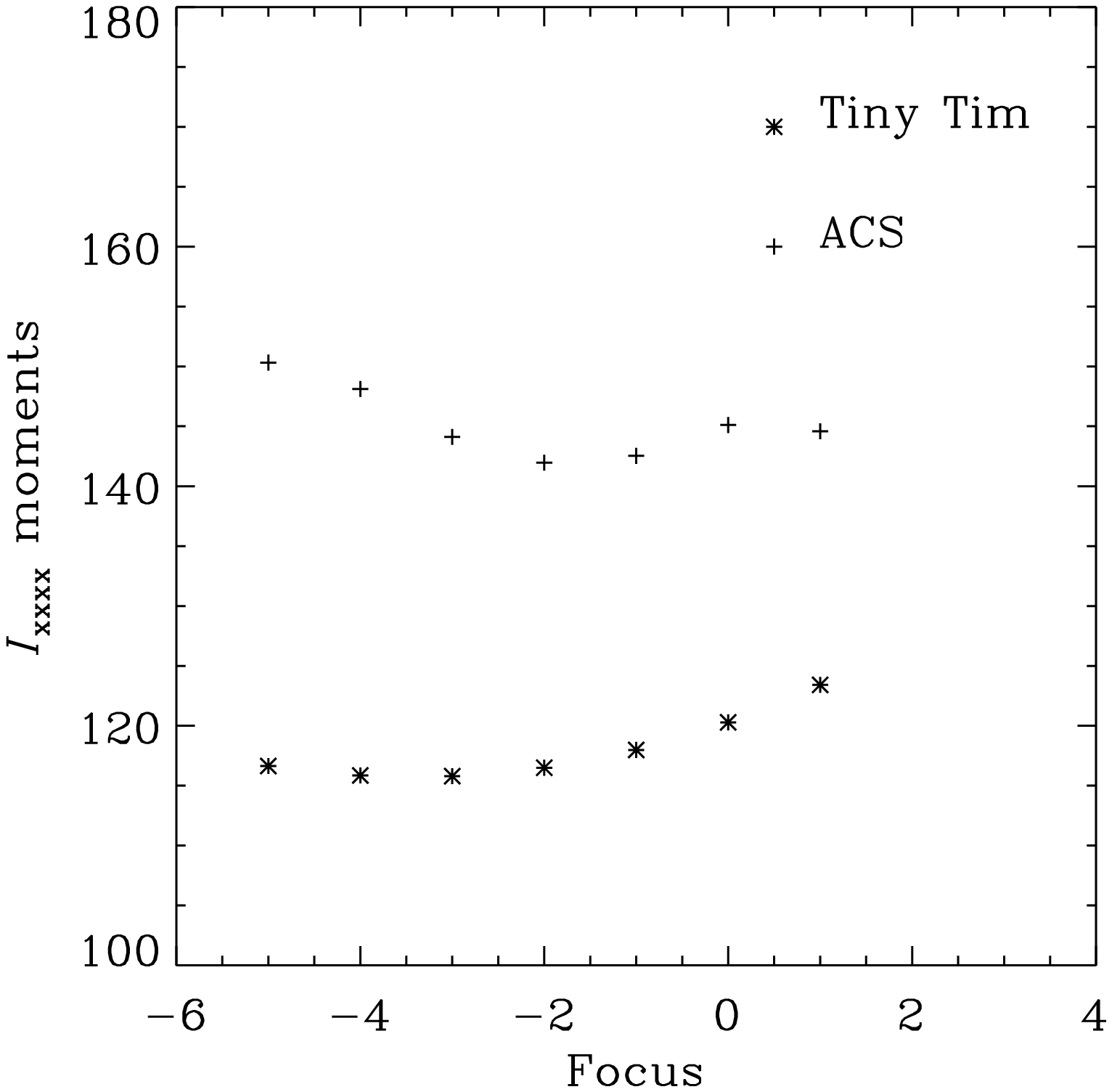}
\epsscale{1}
\caption{ The left panel shows the values of the $I_{xx}$ moments for COSMOS stars and \tinytim\ model stars as a function of telescope focus values.  there is good agreement between the two.  However, the right hand panel shows that the \tinytim\ models are not perfect. The fourth order moment $I_{xxxx}$ is consistently underestimated by \tinytim.
The correction factors between the real and \tinytim\ moments are tabulated in Table~\ref{tab:fudge}.}
\label{fig:xx_and_xxxx}
\end{figure}

For each COSMOS field, we choose the best fit focus value for the telescope as described above.   For each galaxy in that field we correct the measured galaxy moments for PSF effects with the \tinytim\ model star at that focus which is closest to the galaxy's position.  We have obviated the need to interpolate the PSF across the field by creating sufficiently dense grids of model PSFs.  We have found that the \tinytim\ models match the ellipticites (and  more importantly the second order moments) of the COSMOS stars they are meant to represent.  However, the \tinytim\ models do not have the exact profile of the real COSMOS stars and thus the fourth order moments are somewhat too small in the \tinytim\ models. Only the moments with even powers of both $x$ and $y$ are affected. We show, for a range of focus values, the average of $I_{xx}$ and the average of $I_{xxxx}$ for both the \tinytim\ models and the COSMOS stars in Figure~\ref{fig:xx_and_xxxx}. The fourth order moments are used only as a perturbative correction to the second order moments in RRG,
so this slight discrepancy between model and real stars does not significantly effect our PSF correction.   We verified this by multiplying the  \tinytim\ fourth  order moments by a correction factor and re-running our PSF correction routines and found no difference in residual PSF systematics.   For completeness, we tabulate  the multiplicative factor between the \tinytim\ and COSMOS stars in  Table~\ref{tab:fudge}.
It is unclear why the \tinytim\ models have fourth order moments that are slightly too small, but it may have to do with how the charge diffusion kernel is applied.


\begin{table}[h]
 \begin{center}
{\label{tab:fudge}}
\begin{tabular}{|c|c|}
\tableline
Moment & Conversion Factor \\
\tableline\tableline
$I_{xx}$  &  1 \\
$I_{yy}$  &  1 \\
$I_{xy}$  &  1 \\
$I_{xxxx}$  &  1.2 \\
$I_{xxxy}$  &  1 \\
$I_{xxyy}$  &  1.1 \\
$I_{xyyy}$  &  1 \\
$I_{yyyy}$  &  1.2 \\
\tableline
\end{tabular}
\caption{Multiplicative factor between the size of moments in the \tinytim\ models 
and measured values of the COSMOS stars. This is the factor that the \tinytim\ models would have to be multiplied by to have them equal in size to the moments of the COSMOS stars.}
\end{center}
\end{table}


\subsection{Correction for Charge Transfer Efficiency (CTE) Degradation}\label{cte}
Gradual damage  to CCD detectors
due to exposure to charged particle radiation in the harsh environment of space results in a degradation in the efficiency of charge transfer in the CCDs.
High energy charged particle hits create charge
traps that accumulate in the silicon substrate. These traps capture electrons for a short time, and
release them after a time delay drawn from an exponential distribution. Several species of traps can
exist, each with their own exponential time constant. When an image is read out after exposure, the
electrons from sources are stepped across these charge traps. Their capture and subsequent release
trails the electrons away from the original sources, across the image in the opposite direction to
the readout. This spreading of the charge in the readout direction creates problems for photometry,
astrometry, and galaxy shape measurement. Significant effort has already gone into understanding
the CTE of the ACS WFC (Mutchler \& Sirianni
2005) and correcting photometric errors due to charge loss (Riess \& Mack 2004). However, the
subtle changes in the shapes of objects due to CTE degradation have not yet been adequately addressed.

Since ACS was installed in 2002, charge traps have accumulated to the level of
tens per pixel.  There are at least three different trap species, each with different release times for trapped electrons (Sirianni 2006). On the two-CCD WFC, charge readout occurs at the top of the top chip and the
bottom of the bottom chip; charge is read out from the center of the field to the top and bottom
edges. Charge incident furthest from the readout registers is transferred over more pixels and  therefore interacts with more traps during
its translation to the chip edge. The effect on the shape of any one galaxy is
difficult to predict, as it non-linearly depends upon the galaxy's  magnitude, size, ellipticity
and radial profile, as well as its position on the CCD. Furthermore, the fixed number density
of charge traps has the  consequence of affecting faint sources more than bright ones; a
hundred delayed electrons are significant in a source containing a thousand electrons, but not one
containing 100,000. Thus, the shapes of distant galaxies are smeared by this effect, but less so the
bright stars that are typically used to calibrate and test the galaxy shape measurement algorithms
necessary for weak lensing. Thus, CTE degradation  is a particularly difficult  systematic effect to correct,
because it cannot be calibrated using bright, high S/N stars as is typically done with other systematics that affect weak lensing measurements.

\begin{figure}[tb]
\plotone{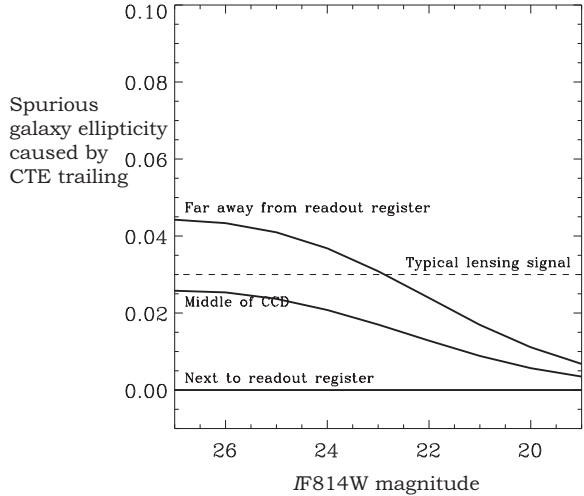}
\caption{An estimate of the spurious ellipticity induced by CTE charge trailing
in a barely resolved, circular galaxy as a function of galaxy flux and at
various positions on the CCD. In practice, the actual amount of spurious
ellipticity also depends upon the intrinsic ellipticity and the  the radial profile of the object. }
\label{fig:curves}
\end{figure}

\begin{figure}[h]
\plotone{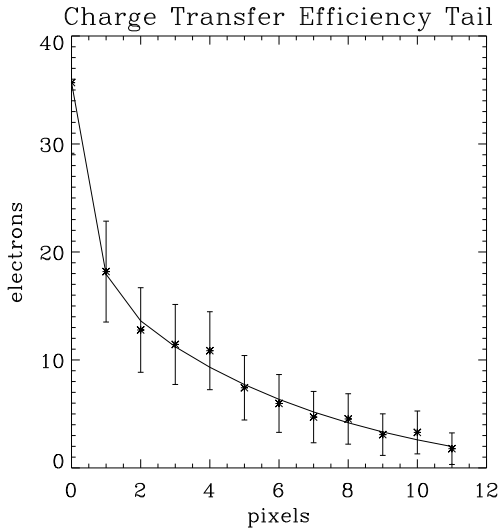}
\caption{Exponential profile created during charge readout as measured in an EPER image as described in \S~\ref{data}.
The value of the pixel that this tail corresponds to is about 2500 electrons.
}
\label{fig:eper}
\end{figure}

\begin{figure}[h]
\plotone{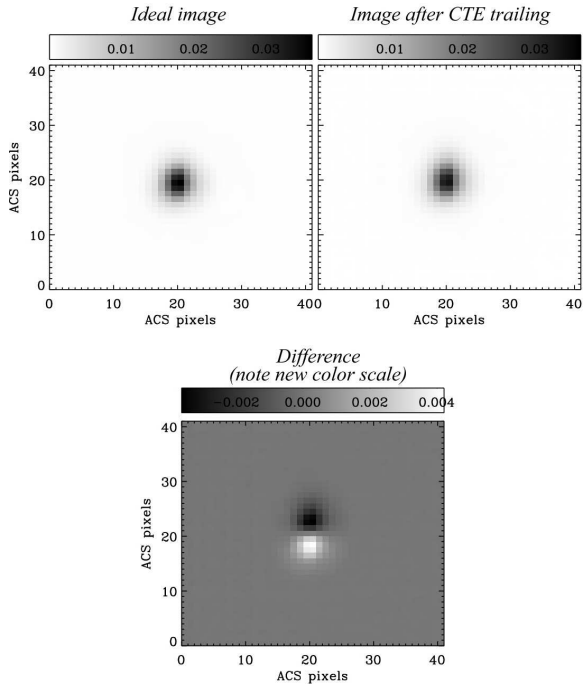}
\caption{  Illustration of the effect of charge trailing on a 25th magnitude
star in single-orbit F814W data, using a crude pixel-level CTE model we have developed. The
readout direction is down.
A simulated galaxy is shown in the left panel as it ought to appear and,
in the right panel, after being read out from the furthest side of the CCD. The difference between the images is small and cannot be detected by eye in the images shown. The difference is, however, significant at the level of precision needed for weak lensing. This
difference  is shown in the bottom panel (with a greatly
expanded grey scale). We eventually hope to understand how charge is transferred between individual pixels in the ACS CCD.  For this work we  correct for the overall effect the CTE has on galaxy ellipticities with Equation~\ref{eq:cte}.
}
\label{fig:images}
\end{figure}

The effect of the CTE degradation is a smearing of objects in the readout ($y$) direction.  For faint galaxies, the size of this effect is comparable to the size of the weak lensing signal we are trying to detect, as shown in Figure~\ref{fig:curves}.  The smearing is caused by  each pixel leaving an exponential trail during readout as shown in Figure~\ref{fig:eper}.  We have developed preliminary models for what CTE degradation does to the charge in individual pixels and we use that to show the effect of CTE degradation on the image of a faint galaxy in Figure~\ref{fig:images}.


\thispagestyle{empty}
\setlength{\voffset}{-20mm}
\setlength{\voffset}{0mm}

Previous weak lensing work with HST has encountered similar problems.  Rhodes \etal\ (2004) corrected for CTE in STIS by creating S/N dependent PSF models.  That is impractical here because we are already dealing with focus-dependent PSF models  and the computing power and complexity required to  model stars across the ACS WFC field at a range of focus values and S/N values would quickly become prohibitive. Furthermore, the CTE degradation grows worse with time and the COSMOS images are taken over a relatively long time period.
Ideally, we would correct each pixel of each image for the effects of CTE degradation as the first step in the image reduction pipeline (see for example Bristow \etal 2002).  CTE effects are the last to go into the  image since they occur during the readout.  However, this would require an exact knowledge of the number of charge traps, the number of species of charge traps (there are at least 3 species each with different release times), and accurate knowledge of the release times of those charge traps.  We are working on a general solution to this problem that incorporates all of that knowledge.  In the meantime, we have developed a parametric equation that allows us to correct the measured ellipticities of galaxies for the effects of CTE degradation.  The CTE depends on the position of the object within  the CCD (electrons further from the readout registers encounter more charge traps), the flux of an object (high flux objects fill the charge traps and the \emph{relative} loss of flux is less), and the date of observation (the CTE is continually degrading due to cosmic ray damage). By assuming that the PSf corrected ellipticities of all the galaxies in the COSMOS fields average to zero, Massey \etal 2006a have found empirically the dependence of CTE effects on these three variables and we use this empirical knowledge to derive the parametric correction equation

\begin{equation}
e_1^{c}=e_1^{M}-e_{\textrm{cte}}
\label{delta_e}
\end{equation}

\noindent where $e_1^{c}$ is the CTE corrected first ellipticity component in the undistorted image coordinate system (which is changed by only a few degrees to the distorted coordinates), and $e_1^{M}$ is the measured first ellipticity component after correction for other PSF affects using the RRG method, as described above. Only the first ellipticity component is affected because this component represents elongation in the $x$ and $y$  direction. $e_2$  represents elongation along axes at $\pm 45 \deg$ to the $x$ axis. The additive correction parameter  $e_{\textrm{cte}}$ is given by

\begin{equation}
e_{\textrm{cte}} = -3\times10^{-5} \times {\rm S/N}^{-1}
                    \times \frac{n_{\rm transfers}}{2048}
                    \times (MJD-52333)
                    \label{eq:cte}
\end{equation}

\noindent where S/N is the object's detection signal-to-noise, $n_{\rm transfers}$ is the distance to the nearest readout register
in native ACS pixels, and MJD is the modified Julian date of the observation. Note that this empirically-derived correction depends on date, $y$ position (number of readouts), and flux, just as the CTE effect is known to.  We were able to obtain similar results by adjusting the exponent on the S/N term while adding a size-dependent term.  However, we choose this formalism because it is simpler and more physically motivated. It is important to note that this particular formalism for the CTE is only valid on data taken before July 2006.  At this time the operating temperature of the ACS WFC CCDs was changed, thus changing the CTE of the CCDs (Sirianni, Gilliand, \& Sembach 2006).

\begin{figure}[h]
\epsscale{1}
\plotone{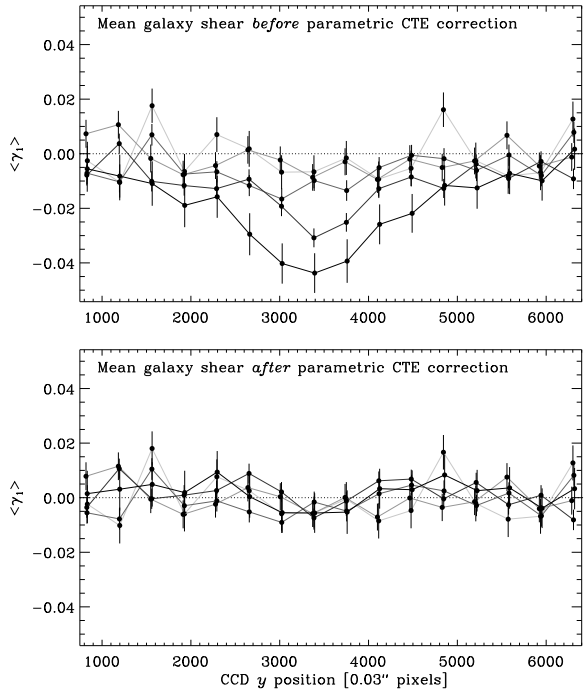}
\epsscale{1}
\caption{The top panel shows the average ellipticity as a function of $y$ position for several magnitude ranges after the RRG PSF correction scheme. The lowest line in the top plot represents galaxies in the magnitude range 26--27. The other lines represent one magnitude bins going up to 22-23. The effects of CTE are clearly shown by the fact that fainter objects (with lower flux) are preferentially elongated in the $y$ direction, corresponding to negative $e_1$.  The problem is worst near the center of the field, farthest from the readout registers. The bottom panel shows that a simple additive factor (as given by Equation~\ref{eq:cte}) on the ellipticity of each galaxy removes the CTE-induced ellipticity.  In the future we hope to understand the CTE well enough to remove CTE effects from the images before a lensing analysis.}
\label{fig:cte_corr}
\end{figure}

We show in Figure~\ref{fig:cte_corr} the ellipticities of galaxies as a function $y$ position and magnitude both before and after the CTE correction for the entire COSMOS survey. The tell-tale dip in $e^M_1$, implying an elongation in the $y$ direction, found in the data is removed by this parametric CTE correction. Before correction, the mean ellipticity of the whole galaxy population is $\langle e^M_1\rangle=-0.020\pm 0.001$, and the gradient of a fit to the faintest magnitude bin in Figure~\ref{fig:cte_corr} is $\frac{\partial e_1}{\partial n_{\rm transfers}}=(-1.74\pm 1.3)\times 10^{-4}$. After correction, these values are reduced to $\langle e_1^{c}\rangle=0.004\pm0.002$ and
$\frac{\partial e_1}{\partial n_{\rm transfers}}=(-1.53\pm 1.0)\times 10^{-5}$. Note that in two-point shear correlation functions (e.g. Massey \etal 2006a) this residual enters only as the value squared. In the sense of that statistic, we have therefore lessened the impact of CTE trailing in faint galaxies by more than two orders of magnitude.

Despite its apparent success, we stress that this prescription is by no means a panacea for CTE effects in
ACS weak lensing data. In this simplified model, we take advantage of the uniform
background level of the COSMOS images to eliminate dependence on this parameter. The other model parameters are also specific to our
dataset and shear measurement method. An improved, pixel-level CTE correction
method, along the lines of Bristow \etal\ (2002) will follow (Massey \& Rhodes, 2007).  This method will take into account the different species of charge traps and their associated release times.


\subsection{Performance of the PSF Correction}\label{performance}

\begin{figure}[tb]
\plotone{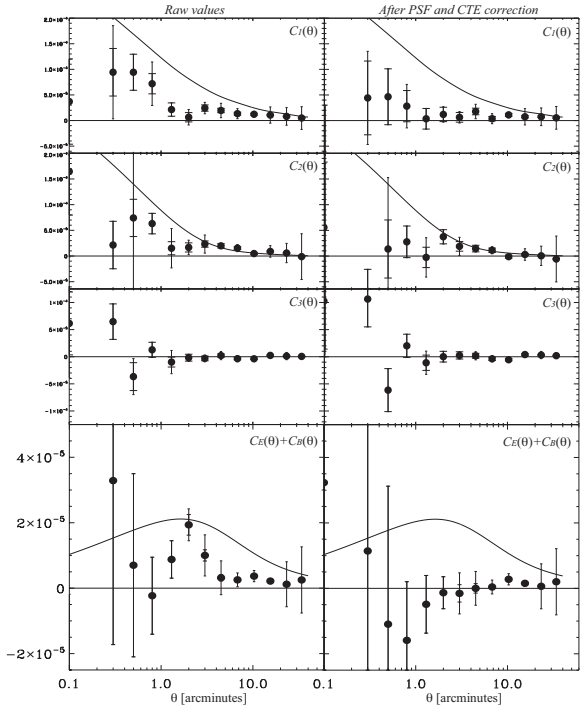}
\caption{Correlation functions between galaxy shears and raw sellar ellipticities, showing the contamination of the shear catalog by residual PSF artifacts. The inner error bars contain statistical errors only and the outer error bars include the variation found by dividing the COSMOS field into four quadrants and performing the correlation function analyses on each quadrant independently.  The left hand panels show the shear measured straight from the images; the right hand panels show the shear after correction. The solid lines show the best-fitting cosmic shear signal measured by Massey \etal\ (2006a) for comparison. The success of our correction scheme is demonstrated by the lowering of the data points to be consistent with zero after correction, especially in the plot of E and B mode signal.}
\label{fig:stargal}
\end{figure}

To demonstrate the correction of the ACS data for PSF and CTE degradation, we show correlation functions between the galaxy shears and the raw stellar ellipticities in Figure~\ref{fig:stargal}. Stars and galaxies should be correlated before PSF correction (due to convolution by the same PSF) but should be uncorrelated after PSF correction.  note that in the RRG method we do not correct the stars for PSF convolution.  Thus, after PSF correction the stars should contain the PSF signal but the galaxies should not, and their ellipticities should be uncorrelated.  Further descriptions of the use of correlation functions in weak lensing can be found in, for instance, Bacon \etal 2003  and Kamionkowski \etal 1998. This figure shows
\begin{eqnarray}
C_{1}(\theta) & = & \frac{\sigma_\gamma}{\sigma_{e^\star}}~
  \big\langle ~ e_{1}^{\star r}({\bf r}) ~
            \gamma_{1}^r({\bf r}+{\bf\theta}) ~
            \big\rangle ~,
\label{eqn:cth_c1} \\
C_{2}(\theta) & = & \frac{\sigma_\gamma}{\sigma_{e^\star}}~
  \big\langle ~ e_{2}^{\star r}({\bf r}) ~
            \gamma_{2}^r({\bf r}+{\bf\theta}) ~
            \big\rangle   \textrm{ and}  \\
\label{eqn:cth_c2}
C_{3}(\theta) & = & \frac{\sigma_\gamma}{\sigma_{e^\star}}\Bigg\{~
  \big\langle ~ e_{1}^{\star r}({\bf r}) ~
            \gamma_{2}^r({\bf r}+{\bf \theta}) ~
            \big\rangle ~+~ \nonumber \\
& & ~~~~~~~~~
  \big\langle ~ e_{2}^{\star r}({\bf r}) ~
            \gamma_{1}^r({\bf r}+{\bf \theta}) ~
            \big\rangle \Bigg\}~,
\end{eqnarray}
\noindent where $\theta$ is the separation between a star and the galaxies, averaging is performed over the whole population and the
superscript~$^r$ denotes components of ellipticity, and shear rotated so that $\hat\gamma_1^r$
($\hat\gamma_2^r$) in each galaxy points along (at 45$\degr$ from) the vector
between the pair. The normalization via star-star correlation functions suggested by Bacon et al.\ (2003) is impractical as a denominator in this case, because the specific PSF pattern of ACS makes it cross zero several times. Figure~\ref{fig:stargal} also shows the star-galaxy correlation functions after separation into E+B modes via the variance of the aperture mass statistic, as defined in Schneider, van Waerbeke \& Mellier (2002).  This figure includes error bars from statistics alone (inner) and, when applicable, variation found be subdividing the COSMOS field into four independent quadrants and doing the analysis on each quadrant independently (outer).  These error bars increase in size after the PSF correction. This is due to the fact that PSF convolution circularizes objects (making their ellipticities smaller).  After PSF correction, the ellipticity distribution of objects is larger and thus the scatter in the correlation functions is larger.

Before correction, the shear measurements contain artifacts from the PSF anisotropy. These are largely removed by the processes described in this section, and are consistent with zero on all scales after correction.  Correlation function $C_1$ and $C_2$ show significant improvement (movement towards zero correlation), especially on scales less than one arcminute. $C_3$ was already nearly consistent with zero before PSF correction and the PSF correction has introduced a larger scatter due to the widening of the ellipticity distribution during PSF deconvolution discussed in the previous paragraph.  The most important test of our PSF correction is shown in the E+B mode plots, which show that after PSF correction the E+B mode is consistent with zero for all scales.  See Massey \etal 2006a for a more detailed description of these correlation functions and the separation of the signal into E+B modes.

\thispagestyle{empty}
\setlength{\voffset}{-25mm}
\setlength{\voffset}{0mm}

\section{Conclusions}\label{conclusions}

We have shown that aliasing of the ACS WFC PSF can be minimized by carefully choosing the parameters of the image reduction pipeline element \multidrizzle.  This aliasing is minimal when \multidrizzle\ is run with a Gaussian kernel, the area of the input pixels containing  the flux is shrunk by a factor of  \texttt{pixfrac}$=0.8$, and the output image has a pixel scale of 0.03 arcseconds per pixel.  We show that the ACS WFC PSF is temporally unstable over the time scale of the COSMOS observations due to thermally induced changes in the telescope's focus.  Using a modified version  of the \tinytim\ software package we create dense PSF grids at a range of telescope focus values from $-10\mu$m to $+5\mu$m. Using the $\sim 10$ suitable stars in each image as taken from the COSMOS lensing catalog (Leauthaud \etal\ 2006), we can calculate the focus value of HST for each COSMOS field within about $\pm 1\mu$m.   We can use the \tinytim\ models to correct the galaxy shapes for the effects of PSF.  There is a residual systematic left after the PSF correction that is due to degradation of the CTE by cosmic ray damage to the CCDs.  We present a parametric correction for the effects of CTE degradation on galaxy ellipticities.  The resulting PSF and CTE corrected catalog has been used to measure cosmic shear by Massey \etal\ (2006).

Our wrappers and implementation of the \tinytim\ code are publicly available from \\ {\url http://www.astro.caltech.edu/$\sim$rjm/acs}.
Since our PSF modelling routines are general, they are applicable to other filter sets and can be used for weak lensing data sets taken with other filters  (e.g. the ACS parallel survey taken with the F775W filter or a re-reduction of the GEMS survey which was taken with the F606W filter). Similarly, the PSF models are not specific to the RRG weak lensing method.  Since we are creating simulated stars, any weak lensing pipeline can make use of these stars for PSF deconvolution. The code and models have already been put to use in other science papers by the COSMOS collaboration in such diverse areas as AGN studies (Gabor \etal\ 2006) and morphological classification (Scarlata \etal\ 2006).

Future work will include a more general correction for the effects of CTE
degradation.  This correction will likely take place on the images themselves as
a first step in the image reduction pipeline. Given the general nature of our
PSF models, we plan to use them to perform further weak lensing analyses on the
COSMOS data set with the more advanced ``shapelets'' method
(Refregier 2003; Refregier \& Bacon 2003; Massey \& Refregier 2005).

 \acknowledgments
We thank Stefano Casertano and Andy Fruchter for useful discussions on the ACS PSF and how it is affected by drizzle.
Our use of the \tinytim\ software would not have been possible without the help of John Krist.
 The HST COSMOS Treasury program was supported through NASA grant
HST-GO-09822. We wish to thank Tony Roman, Denise Taylor, and David
 Soderblom for their assistance in planning and scheduling of the extensive COSMOS
 observations.
 We gratefully acknowledge the contributions of the entire COSMOS collaboration
 consisting of more than 70 scientists and to the COSMOS PI Nick Scoville in particular.
 More information on the COSMOS survey is available \\ at
  {\bf \url{http://www.astro.caltech.edu/~cosmos}}. It is a pleasure the
 acknowledge the excellent services provided by the NASA IPAC/IRSA
 staff (Anastasia Laity, Anastasia Alexov, Bruce Berriman and John Good)
 in providing online archive and server capabilities for the COSMOS datasets.
 The COSMOS Science meeting in May 2005 was supported in part by
 the NSF through grant OISE-0456439.  We thank the anonymous referee for useful comments that improved the paper immensely. We also thank Matthew Lallo and Russell Makidon for providing focus data for the HST during the time the COSMOS observations were made. RM was supported in part by grant HST-AR 10964.

 {\it Facilities:} \facility{HST (ACS)}.


 \end{document}